\documentclass[onecolumn,usenatbib]{mnras}
\usepackage{graphicx}
\usepackage[export]{adjustbox}
\usepackage{xcolor}
\newcommand{\Dpar}{D_{\parallel}}
\newcommand{\Dperp}{D_{\perp}}

\newcommand{\lskip}{\vskip \baselineskip}
\newcommand{\nskip}{\lskip \noindent}
\newcommand{\bm}[1]{\mbox{\boldmath$ #1 $}}
\newcommand{\grad}{\mbox {\boldmath $\nabla$}}
\newcommand{\bdot}{\mbox{$\bm{\: \cdot \:}$}}
\newcommand{\halfskip}{\vskip 0.5\baselineskip}
\newcommand{\be}{\halfskip \begin{equation}}
\newcommand{\ee}{\end{equation} \halfskip \noindent}
\newcommand{\ba}{\halfskip \begin{eqnarray}}
\newcommand{\ea}{\end{eqnarray} \halfskip \noindent}

\newcommand{\hatz}{\bm{\hat{z}}}

\newcommand{\hatb}{\bm{\hat{b}}}
\newcommand{\ncr}{n_{\rm cr}(\varpi \: , \: z \: , \: t)}
\newcommand{\ncrs}{\tilde{n}_{\rm cr}(\varpi \: , \: z \: , \: s)}
\newcommand{\Hcr}{H_{\rm cr}}
\newcommand{\half}{\mbox{$\frac{1}{2}$}}
\newcommand{\Msun}{M_{\odot}}

\begin{document}
\title[The contribution of SNRs on the CR flux at Earth]
{The contribution of  nearby supernova remnants  on the cosmic ray flux at Earth  }
\author[AL-Zetoun, Achterberg]{A. AL-Zetoun$^1$\thanks{E-mail: a.al-zetoun@astro.ru.nl}  ,     A. Achterberg \\  
$^1$Department of Astrophysics, IMAPP, Radboud University, Nijmegen, P.O. Box 9010, 6500 GL Nijmegen, The Netherlands}

\date{Accepted..., Received...; in original form ...}
\pubyear{2018}

\maketitle
\begin{abstract}
{ 

We consider anisotropic diffusion of Galactic cosmic rays in the Galactic magnetic field, using the Jansson-Farrar model for the field.  
In this paper we investigate the influence of source position on the cosmic ray flux at Earth in two ways: 
[1] by considering the contribution from cosmic ray sources located in different intervals in Galacto-centric radius, and [2] 
by considering the contribution from a number of specific and  individual close-by supernova remnants. 
Our calculation is performed by using a fully three-dimensional stochastic method. This method is based on the numerical solution of a set of 
stochastic differential equations, equivalent to It\^o formulation, that describes
the propagation of the Galactic cosmic rays. 
 
}
\end{abstract}

\begin{keywords}
Methods: numerical -- diffusion -- magnetic fields -- cosmic rays--supernova remnants
\end{keywords}

\section{Introduction}
\label{intro:1}

Galactic cosmic ray (CR) sources, presumably core-collapse type II supernovae, are thought to be concentrated in the Galactic disk. 
Therefore, the effective source density should have a radial distribution  similar to that of  type II (core-collapse) supernova remnants (SNRs), 
see for instance \citet{StrongM2007}. It is generally believed that 
these SNRs  are thought to be the {\em main} source of Galactic CRs (GCRs), see for instance \citet{Blandford1987}.
Additionally, the strong winds from O-B stars 
(e.g. \citet{Webbetal} and \citet{ThouRach}), neutron stars, see for instance \citet{Venkatesan}
and pulsar winds may very well contribute to the observed CR flux at Earth. In this study we will disregard these possible CR sources, concentrating
on CRs from core-collapse supernovae.
As a consequence, in our model the distribution  from Type II supernovae determines the CR source distribution. 
The distribution of Galactic SNRs  derived in the literature is poorly constrained for large Galacto-centric radii.
Observations have found that the distribution of SNRs extends approximately up to a radial distance of $R \le \; 16\rm \; kpc $, 
assuming that the SNR distribution is {\it grosso mode} axi-symmetric with respect to the Galactic Center \citet{CaseB1996}. 

\nskip
In a simple {\em Leaky Box Model} , where the typical CR residence time in the Galaxy is determined by the height $H_{\rm cr} \simeq 4-10 \; \rm  kpc$ 
of the CR halo above the Galactic disk and by
the typical  CR diffusion coefficient $D_{\rm cr}$ as $t_{\rm res} \simeq H_{\rm cr}^2/2 D_{\rm cr}$, 
the contribution to the observed CR flux from sources at a distance from the observer much 
larger than $H_{\rm cr}$  rapidly becomes negligible.  
For instance, in the case of isotropic diffusion with instantaneous escape at $z = \pm H_{\rm cr}$, the CR density 
in the Galactic mid-plane due to a source at a give distance $\varpi$ decays for $\varpi \gg \Hcr $ as (see Appendix):
\be
	 n_{\rm cr}(\varpi \: , \: 0 \: , \: t= \frac{\varpi \Hcr}{\pi D_{\rm cr}} \: \tau) \propto 
	 \frac{\pi D_{\rm cr}}{\varpi \Hcr \: \tau} \: \displaystyle {\rm exp}  \left(- \frac{ \pi \varpi}{4 \Hcr} \: \tau \right) \; .
\ee
It is assumed that CRs are injected impulsively at $t = 0$, see the Appendix for details. 
Here the dimensionless time $\tau$ is given by $\tau \equiv \pi \: D_{\rm cr} \: t/\varpi \Hcr$, 
with $\tau \simeq 1$ corresponding 
to moment of the the maximum CR density at given $\varpi \gg \Hcr$ as a function of time.
\nskip
The effect of nearby discrete  sources  to the observed CR nuclei has been studied extensively in a number of different models. 
For instance \citet{Erlykin2006}, studied the anisotropy observed at  Earth from local sources located within a distance of few ${\rm kpc}$ from the Sun. 
They considered sources that are distributed randomly in the nearby Galaxy. They concluded that the anisotropy of arrival directions is produced mainly by these local sources. \citet{Strong2001} used the {\small GALPROP}  code to  study the effect of the discrete sources of SNRs on the CR  proton densities in the Galaxy. 
\citet{Thoudam2007} studied the CR density fluctuations  from a known nearby source located within a distance of $ 1.5 \; {\rm kpc}$ from Earth, and the influence of the nearby SNRs  on the observed CR anisotropy below the knee in the CR spectrum. 
\nskip
The propagation of the GCRs  from their source to the observer is usually described in the diffusion approximation. In general, the diffusion of charged CRs is thought to be due to the deflection by stochastic Lorentz forces, either due to irregularities in the Galactic magnetic field, which is known to have a strong stochastic component,  or due to self-excited Alfv\'en waves, see for instance \citet{Berezinskii1990}  and \citet{StrongM2007}. 
\nskip
Several quasi-analytical models have been proposed to describe CR propagation in the Galaxy as a whole. The simplest such model is the
{\em Leaky Box Model} that considers the {\em global} effects of CR propagation inside the Galaxy, and escape from the Galaxy. 
It was proposed by \citet{ Cowsik1973} as an improvement on the simpler slab model \citet{Lezniak1976} that was in use at that time. The {\em Leaky Box Model}  considers the Galaxy as a box with transparent walls around the Galactic disk.
The CR sources are distributed uniformly in the Galactic  disk, and the CRs diffuse within the Box, escaping as they reach it is  upper and  lower edge.
Generally speaking, the {\em Leaky  Box Model}  satisfactorily  reproduces existing data on stable primary and secondary nuclei in CRs,
but with some limitations, see for instance \citet{StrongM2007}.  
\nskip
Nowadays, various widely-used numerical codes are available that concurrently calculate 
the propagation and CR interactions (such as losses, spallation in the interstellar gas, re-acceleration etc.) 
of the Galactic CRs in the Galaxy. Examples of such codes are
{\em GALPROP}, see \citet{StrongM2010}, {\em DRAGON}, see \citet{Evoli2008} and {\em CRPropa 3.1} , see \citet{Merten2017}, as well as {\em PICARD}  \citet{Kissmann}.
 {\small GALPROP} code is the most widely code and has been developed to overcome the limitations of analytical and semi-analytical models like the Leaky Box model. 
The code calculates the transport and interactions of the propagation of CR nuclei, electrons and positrons, anti-protons. 
However, it has not yet  implemented a method for dealing with anisotropic diffusion. {\small  DRAGON} code is  a numerical tool designed to cover all relevant processes include GCRs and their secondary products over a very wide energy range, that addresses for example the anisotropic diffusion problem. 
The recent {\small  CRPropa} update {\small  CRPropa 3.1} also solves the CR spatial transport equation using stochastic differential equations, 
the method also used in this paper. This is the most straightforward way  
to treat  anisotropic  CR diffusion with respect to an arbitrary magnetic field. 
Finally, the {\small  PICARD} code is a more sophisticated numerical tool that uses advanced contemporary numerical algorithms for solving the 
 diffusion equation that compute a steady state solution for the CRs propagation problem without any modification of numerical parameters (such as energy dependence of
 diffusion coefficients or parameters associated with the nuclear network that is used to calculate CR spallation) by the user.
\nskip
The outline of this paper is as follows: In Section \ref{sec:2}, we present the propagation model and the relevant assumptions, like the source distribution of CR, 
the diffusion tensor and it is connection to the large scale Galactic magnetic field. 
In Section \ref{sec:3} we present the spatial distribution of CRs from  a uniform source as well as from known SNRs, and discuss the results. 
We do this for two values for the ratio of the parallel - and perpendicular diffusion coefficients: $\Dperp/\Dpar = 1$ (isotropic diffusion) and $\Dperp/\Dpar = 0.01$
(strongly anisotropic diffusion). In Section \ref{sec:4} we present the conclusions.

\section{ Model description}
\label{sec:2}
Recently,  \citet{ALZetoun2018} have proposed a fully three-dimensional model  for CR propagation in the Galactic magnetic field,
based on the numerical solution of a set of 
stochastic differential equations in the It\^o formulation.
We solve these equations using a fully-anisotropic diffusion tensor, the situation that is expected in the presence of a large-scale ordered Galactic magnetic field.
\nskip
The CRs are represented by test particles following a trajectory $\bm{x}(t)$ that change their position by an amount $\Delta \bm{x}$ in a time-span $\Delta t$ 
according to the prescription:
\be
\label{TPmotion}
	\Delta \bm{x} = \bm{V} \: \Delta t + \Delta \bm{x}_{\rm diff} \; .
\ee
The vector $\bm{V}$ is a mean velocity, formally defined as:
\be
\label{Vdef}
	\bm{V} = \bm{U} + \grad \bdot \bm{\mathrm{D}} \; ,
\ee
in component notation $V_{i} = U_{i} + \partial D_{ji}/\partial x_{j}$, 
with $\bm{U}$ the bulk velocity of the (ionized) interstellar medium and $\bm{\mathrm{D}}$ the CR diffusion tensor. 
For anisotropic diffusion with respect to the large-scale field (see Section \ref{sec:2.2} for more details), with a diffusion tensor $\bm{\mathrm{D}} = \Dpar \: \hatb \hatb +
\Dperp \: (\bm{\mathrm{I}} - \hatb \hatb)$ where $\hatb$ is the unit vector along the magnetic field and $\bm{\mathrm{I}} = {\rm diag}(1 \: , \: 1 \: , \: 1)$ is the unit tensor,
the diffusive step $\Delta \bm{x}_{\rm diff}$ equals in the It\^o formulation:
\be
\label{diffstepsize}
	\Delta \bm{x}_{\rm diff} =    \sqrt{2 \Dperp \: \Delta t}\; \xi_{1}\: \hat{\bm{e}}_{1}
	+  \sqrt{2 \Dperp \: \Delta t}\: \xi_{2} \;	\hat{\bm{e}}_{2} + \sqrt{2 \Dpar \:  \Delta t}\; \xi_{3} \: \hat{\bm{e}}_{3} \; .
\ee
Here we employ local Cartesian coordinate system with base vectors  $\hat{\bm{e}}_{1} \: , \: \hat{\bm{e}}_{2} \: , \: \hat{\bm{e}}_{3}$ with the local large-scale 
magnetic field along $\hat{\bm{e}}_{3} = \hatb$. 
In recipe (\ref{diffstepsize}) the quantities $\xi_{1}$, $\xi_{2}$ and $\xi_{3}$ are three independent (that is: statistically {\em un}correlated) 
stochastic variables, with zero mean and unit dispersion, i.e. in simplified notation: $< \xi > = 0$ and $<\xi^2> = 1$.
For more details about the model see \citet{ALZetoun2018}. Note that $\sqrt{2 D_{\perp} \: \Delta t}$ and $\sqrt{2 D_{\parallel} \: \Delta t}$ are the rms diffusion
distances for a timespan $\Delta t$ in a given direction, either perpendicular to the large-scale magnetic field or along the magnetic field.
\nskip
In what follows we put $\bm{V} = 0$, thereby only including diffusive process and neglecting  the effects  of a possible  Galactic wind, curvature and gradient 
drift of CR guiding centers in the Galactic magnetic field for simplicity. These effects will be considered in a follow-up paper.
We also limit ourselves to a parameter study, focusing on the effect of using different values for the ratio of the perpendicular and parallel diffusion coefficients, $\Dperp/ \Dpar$. We take $\Dperp/\Dpar \equiv \epsilon$ $ \;= 0.01,\; 1.0$. In the last case we formally recover isotropic diffusion in three dimensions.
In the simulation, we remove particles once they reach the upper/lower boundary of the CR halo at $z = \pm \Hcr$, corresponding to free escape.
This escape determines their residence time $t_{\rm res}$ in the Galaxy.
\nskip

\subsection{The distribution of cosmic ray sources}
\label{sec:2.1}

In our previous  work ( \cite{ALZetoun2018}) we described the diffusion of CRs originating from a single point source that was placed
at various locations in the Galactic mid-plane $z = 0$. We injected the particles impulsively (that is: in a burst of negligible duration)  at a 
fixed Galacto-centric radius close to that of the Sun, which is at $R_{\odot} = 8.5\; {\rm kpc}$.
We assumed a uniform diffusion coefficients $\Dperp$ and $ \Dpar$, 
using typical values for $\Dpar$ derived from the (isotropic) Leaky Box Model and the aforementioned simulations for CR energies of $\sim \;  1 \; \rm GeV/nucleon $.
\nskip
In the present paper we investigate the flux near Earth from multiple CR sources. We do this using two different approaches:
\begin{enumerate} 
\item First we consider the flux due to multiple sources in a fixed interval in Galacto-centric radius. In this case CRs originate in a ring in the 
mid-plane of the Galactic disk, centered on the Galactic Center. We do not use discrete source positions, 
but assume a smooth distribution for the source locations in azimuth with respect to the Galactic Center.
We inject the CR particles in the mid-plane $(z=0)$ of the Galactic disk and distribute the source locations uniformly and randomly in 
four rings  around the Galactic Center, with cylindrical injection radius $r_{\rm inj}$ in the four intervals (in kpc)
$2<r_{\rm inj}\le 4$,  $4<r_{\rm inj}\le 6$, $6<r_{\rm inj}\le8$ and $8<r_{\rm inj}\le 10$. 
The distribution in cylindrical radius $r_{\rm inj}$ is obtained 
for each injected particle from the simple prescription: 
\be
\label{uniform}
 r_{\rm inj} = \sqrt {(r^{2}_{\rm max} - r^{2}_{\rm min})\; \Xi + r^{2}_{\rm min}\;} \:  .
\ee
The two radii $r_{\rm min}$ and $r_{\rm max}$ are the inner and outer radius of each ring and $\Xi$ is a random variable distributed uniformly between $0$ and $1$. 
The azimuthal angle $\phi$ of the injection site (where $\phi$ is the cylindrical polar coordinate in the plane of the Galactic Disk with respect of the Galactic Center) 
is chosen randomly from a uniform (top-hat) distribution between $0$ and $2 \pi$.
This prescription guarantees a uniform (projected) surface density of sources in the Galactic plane. In this case all CRs are injected at time $t = 0$ and followed
until they escape the Galaxy as they reach a height $H_{\rm cr}$ above/below the Galactic mid-plane.

\item
We look at the contribution of a number of specific sources close to Earth.
We first consider four hypothetical equidistant sources (distance: 5 kpc, see Figure \ref{figure5} (c)) in order to study how the 
local magnetic field geometry
influences their propagation to Earth. Next we consider CRs from the same set of sources as employed by \citet{Thoudam2007}, 
see  Table \ref{table:1} in Section \ref{sec:3.3}.
These CRs are released from their sources at the estimated time of the associated supernova explosion, see below for a discussion of this assumption.
In the second case we consider only known SNRs located within a distance of $1.5\; {\rm kpc}$ from the Solar System, with an age $\le \: 2.4 \times 10^{4}\; {\rm yr}$. 
This age is much smaller than the typical residence time of CR in the Galaxy, which is $t_{\rm res} \simeq 10^{7} \: {\rm yr}$ in these simulations. 
This means that most of the CRs produced in these sources are still diffusing in the Galaxy.
We note in passing that the residence time is set by the assumed value of $\Dpar$ with $t_{\rm res} \propto \Dpar^{-1}$, and can be re-scaled
to different values by changing that parameter.
 
These local sources are expected to produce temporal significant variations in the CR densities at Earth, see the discussion in \citet{Thoudam2006}. 
Table \ref{table:1} shows the distance of each known SNRs from the Solar System, their  exact position in Cartesian coordinates $(x \: , \: y)$
in the mid-plane centered on the Galactic Center, 
with the Solar System at  $x = - R_{\odot} = - \; 8.5\; {\rm kpc}$, Galactic longitude $l$ , as well as the age as given in \citet{Thoudam2007}. 
More details regarding these sources can be found in the SNR catalog of \citet{Green2017}.

\end{enumerate}
In both cases we define a {\em local volume} around the Sun with a radius of $1\; {\rm kpc}$. We record each time step 
how many CRs reside inside this volume and add the result over the different steps during the entire run of the simulation. 
We return to the reason for  this approach below. 
We use as a criterion for \lq{}residence\lq{} $| \bm{r}_{\rm CR} - \bm{r}_{\odot} | < 1  \;{\rm kpc}$, with $\bm{r}_{\rm cr}$ 
the position of a CR and $\bm{r}_{\odot}$ that of the Sun.
\nskip
The CRs in the vicinity of the Solar System diffuse in a halo around the Galactic disk with height $H_{\rm cr}$. 
The estimated value for $H_{\rm cr}$ lies in the range $(4-10)\; {\rm kpc}$, see for instance  \citet{StrongM2010}. 
CRs escape at the upper (lower) halo boundary, i.e. at $z\;=+ H_{\rm cr}$ ($z=- H_{\rm cr} $). 
The local volume around the Solar System, defined here, is small enough that edge effects due to CR escape play no role, 
except in determining the local CR density. We take the CR density inside this local volume, centered on the position of the Solar System, 
as a measure for the observed CR flux at Earth.

We neglect the effects of the Solar Wind on the observed CR flux. 
We also neglect the effects of nuclear spallation of CRs in the interstellar gas.

\nskip
\subsection{Anisotropic Diffusion  }
\label{sec:2.2}
The presence of a large-scale (regular) magnetic field clearly breaks isotropy as the field introduces a preferred direction. 
This is important for the calculations presented here as this will influence the contribution from different sources to the local CR density. 
Quite generally, one expects diffusion along the field to be more rapid than diffusion across the field. 
As demonstrated in  \citet{ALZetoun2018}, the ratio of the perpendicular and parallel diffusion coefficients, $\epsilon \equiv \Dperp / \Dpar$, see also
Eqn. \ref{difftensor} below, regulates the average residence time for the CR population in the Galaxy in the Jansson-Farrar field (see Section \ref{sec:2.3}) we adopt. 
It increases strongly for small values of $\epsilon$. When considering the local CR flux entering the Solar System one expects that sources 
that are \lq{}magnetically connected\rq{}, that is: sources on or close to those field lines of the regular field that cross the local volume, 
will tend to  have a larger contribution to the local CR density than \lq{}unconnected\rq{} sources.
\nskip
For anisotropic diffusion in a large-scale Galactic magnetic field true diffusion (as opposed to drift) 
is characterized by a diffusion tensor $\bm{\mathrm{D}}$ that needs to take into account of the different values for the perpendicular and parallel diffusion coefficients. 
The diffusion tensor is given by:
\be
\label{difftensor}
	\bm{\mathrm{D}} = D_{\perp} \: \left( \bm{\mathrm{I}} - \hatb \hatb \right) + \Dpar \: \hatb \hatb = \Dpar \: 
	\left[ \: \epsilon \: \left( \bm{\mathrm{I}} - \hatb \hatb \right) +  \: \hatb \hatb \: \right] \; ,
\ee
with $\bm{\mathrm{I}}$ the $3 \times 3$ unit tensor, $\bm{B}$ is the ordered magnetic field, $\hatb \equiv \bm{B}/|  \bm{B} |$  is the unit vector along the magnetic field 
and $\hatb \hatb$ is the associated dyadic tensor with components $(\hatb \hatb)_{ij} = \hat{b}_{i} \: \hat{b}_{j}$.  
In fact, we assume that  the diffusion along the large-scale field (governed by the parallel diffusion coefficient $\Dpar$ ) and the  diffusion across the field 
(governed by the perpendicular diffusion coefficient $\Dperp$ ) may reasonably depend only on the micro-physics of the ( axi-symmetric)  diffusion on in a locally field-aligned coordinate system. 
In this paper we consider spatially uniform diffusion coefficients $\Dperp$ and $\Dpar$, where  $\Dperp$ and $\Dpar$ have an energy dependence of the form that 
usually described by empirical diffusion models, 
see for instance \citet{Berezinskii1990} and \citet{StrongM2007}. 
Typically, for CRs with energy $E$ and charge number $Z$ isotropic diffusion models give
a diffusion coefficient of order
\be
\label{rigidity}
   D(E)= D_{0}\; \bigg( \frac{E/Z}{1 \: {\rm GeV/charge}}\bigg)^{\delta}, 
\ee
\nskip
where $D_{0} = 3 \times 10^{28}$ cm$^{2}$ s$^{-1}$ and $\delta \simeq 1/3$.
\nskip
In our calculation we adopt this scaling for both 
$D_{\parallel}$ and $D_{\perp} = \epsilon \: \Dpar$ since we assume that $\epsilon$ is constant. As long as one considers particles of a single energy/nucleon this
is not really a problem as a different scaling with energy of $\Dperp$ and $\Dpar$ can be absorbed into the value of $\epsilon$. 
\nskip
There is one case in which the assumption
of a constant $\epsilon$ can be justified for different CR energies as $\Dperp \propto \Dpar$. This is the case where the random magnetic field component $\delta \bm{B}_{\perp}$ 
perpendicular to the regular magnetic field has a correlation length smaller than the scattering mean free path $\lambda_{\parallel}$ along the field. 
Note that by definition $\Dpar = \lambda_{\parallel}c/3$  for relativistic CRs.
If one has $\Dperp^{\rm t} \ll \Dpar$ with $\Dperp^{\rm t}$ the \lq{}ordinary\rq{} cross-field diffusion coefficient due to very small-scale turbulence, the tilt of the {\em total} field $\bm{B} = \bm{B}_{0} + \delta \bm{B}_{\perp}$ with respect
to the regular field $\bm{B}_{0}$ leads to an effective cross-field diffusion coefficient (again with respect to the regular field) of order:
\be
	D_{\perp}^{\rm eff} \simeq \Dpar \: \left( \frac{| \delta \bm{B}_{\perp}|}{B_{0}} \right)^2 \; .
\ee
This result holds {\em provided} the correlation length of the random field component along the regular field is sufficiently small, 
see the careful discussion in \citet{Isichenko} for the
conditions needed for the validity of this result.
\nskip
In the calculations presented here we do not assign an energy to the particles, but take typical values for CRs with $E/A \sim 1\; \rm GeV $ per nucleon.

\nskip
Experimental studies of the B/C ratio and the sub-Fe/Fe ratio suggest that $\delta$ changes between ( 0.6 to 0.3) in the ${\rm GeV\; -\; TeV}$  region. 
The value of $D_{0}$ is only weakly constrained by measurements of the observed flux of stable nuclei. 
For example: unstable/stable ratios ($B^{10}/ B^{9}$) need large CR halo $H_{\rm cr} \simeq \; 5 \; {\rm kpc}$, which leads to large values of  
$D_{0}\simeq (3-5) \times 10^{28}$ cm$^{2}$ s$^{-1}$ at $E_{0}=  1\; {\rm GeV}$ per nucleon, see for instance \citet{StrongM2007} and  \citet{Evoli2008}.

\subsection{ Model for the large-scale Galactic magnetic field  }
\label{sec:2.3}
 
We employ the GMF model of \citet{JF2012A}, and \citet{JF2012B} (henceforth: the JF model) to describe the large-scale GMF. 
These authors use a Galacto-centric $(r\:,\: \phi\:,\: z)$, as well as a  
right-handed Cartesian $(x\:,\: y\:,\: z)$  coordinate system, with the Sun located along
 the negative  $x$-axis, at $x_{\odot} =-8.5\; \rm {\rm kpc}$. The Galactic north is in the positive $z$-axis, and $\phi = 0$ is along the positive $x$-axis. 
 The magnetic field  is set to zero for $r > 20\; \rm {\rm kpc}$.
\nskip
The JF model is a three-component model consisting of a disk component, a toroidal component, and the so-called {\em X-field}: 
\begin{enumerate}
\item The disk component is  described by a toroidal field  in the $x-y$  plane that includes  a molecular ring between $3-5 \; {\rm kpc}$, 
which is purely azimuthal with a constant field strength, and a logarithmic spiral arm structures (eight arms in total) at larger radii up to 20 kpc, 
where the magnetic field has a different value in each arm.
\item  The halo component has a purely toroidal (azimuthal) component with different radial and vertical extent
in the northern and southern halo.
 \item  The {\em X-field} component is purely poloidal  and  axi-symmetric, (i.e. no azimuthal component).
Given field lines are straight, with an inclination angle determined solely by the cylindrical radius $r_{\rm p}$ where the field crosses the Galactic mid-plane.
\end{enumerate}
For  all  detailed description of the model and it is parameters  we refer to,  see \citet{JF2012A}, and \citet{JF2012B} and references therein, 
as well as the discussion in Section 2 of  \citet{ALZetoun2018}.
\subsection{Grammage and the density of the diffuse interstellar medium }
\label{sec:2.4}

CRs accumulate grammage  $\Sigma_{\rm cr} \sim ({\rm pathlength}) \times ({\rm mass \; density})$ as they traverse the Galactic disk. 
This grammage is an important quantity as it determines the number of spallation reactions that a typical primary CR undergoes.
In our calculation the grammage is accumulated each time step and given by:

\be
\label{grammage}
	\Delta \Sigma_{\rm cr} = \rho(\bm{r}_{\rm cr} ) \: v \: \Delta t \; ,
\ee
\nskip
where $\rho$ is the density of the diffuse gas,  $v = c \: \sqrt{1 - (mc^2/E)^2}$ is the velocity of a CR with rest mass $m$ and energy $E$ 
and $\bm{r}_{\rm cr}$ is the instantaneous position of the CR inside the Galaxy. 
In our calculation, we keep $E$ constant, and consider particles with $E \gg mc^2$ so that $v \simeq c$.
\nskip
 The density of the diffuse gas in the disk  mid-plane of the Galaxy (i.e. at $z = 0$) from the relative abundance of secondary CRs  scales with radius $r$ as, see for instance \citet{Kalb2009}:
\be
\label{cdens}
	\rho(r \: , \: 0) = \left\{ \begin{array}{ll}
	\rho_{0} & \mbox{for $r < R_{\rm c} = 7$ kpc}  \; , \\
	& \\
	\rho_{0} \: {\rm exp}\left[ \: -(r - R_{\rm c})/R_{\rm d} \: \right] & \mbox{for $r > R_{\rm c} = 7$ kpc}  \; . \\
	\end{array} \right. 
\ee
The scale length in the exponential is $R_{\rm d} \simeq 3.15$ kpc, and $\rho_{0} \simeq 3 \times 10^{-24} \; {\rm  g} \; {\rm cm}^{-3}$. 
The typical thickness $H_{\rm d}$ of the hydrogen disk flares out as $r$ increases: it scales as:
 It is expected that this law will break down for $r < 5$ kpc, but we will use it anyway for lack of a better model.
In view of this we can adopt an axi-symmetric density distribution that scales with height $|z|$ above the mid-plane as:
\be
	\rho(r \: , \: z) = \rho(r \: , \: 0) \: {\rm exp}\left( - \frac{|z|}{H_{\rm d}(r)} \right) \; .
\ee
The typical thickness $H_{\rm d}$ of the hydrogen disk flares out as $r$ increases: it scales as:
\be
	H_{\rm d}(r) = H_{0} \: {\rm exp}(r/R_{\rm h}) \; ,
\ee
with $R_{\rm h} \simeq 9.8$ kpc and $H_{0} \simeq 0.063$ kpc. It is expected that this law will break down for $r < 5$ kpc, but we will use it anyway for lack of a better model.

\section{The simulations }

\label{sec:3}
In this Section we describe the results of the diffusion of CRs origination from multiple sources at different Galacto-centric radii  for two approaches. 
In the first case, we inject the  CR particles uniformly and randomly in the mid-plane of the Galactic disk, at fixed Galacto-centric radius close to the Sun in four rings, 
assuming a smooth distribution of source locations in azimuth with respect to the Galactic center. 
In the second case, we inject the CR particles from discrete sources  listed in table \ref{table:1}  in the mid-plane of the Galactic disk, as well as from discrete sources 
at a certain distance from the Sun. In all cases, particles are injected at time $t =0$ from their sources. We then let the CRs  diffuse according to prescription (\ref{diffstepsize}),
assuming uniform diffusion coefficients $\Dperp$ and $\Dpar$, as well as no mean flow. 
The flux and the accumulated grammage are calculated  in the local volume around the Solar System from  multiple sources at different Galacto-centric radii.
\nskip
We want to emphasize that the distribution of CR ages obtained by this method for the CRs inside the local volume centered on the 
Solar System is {\em not} the escape time from the Galaxy, but the CR age during residence inside the local volume. This is always significantly less than the
escape time. The escape time  $t_{\rm res}$ is the total time that a CR spends in the Galaxy after injection.

\subsection{Results for CRs from multiple sources in four Galacto-centric rings}
\label{sec:3.1}
\begin{figure}
 \centering
\includegraphics [width=\columnwidth]{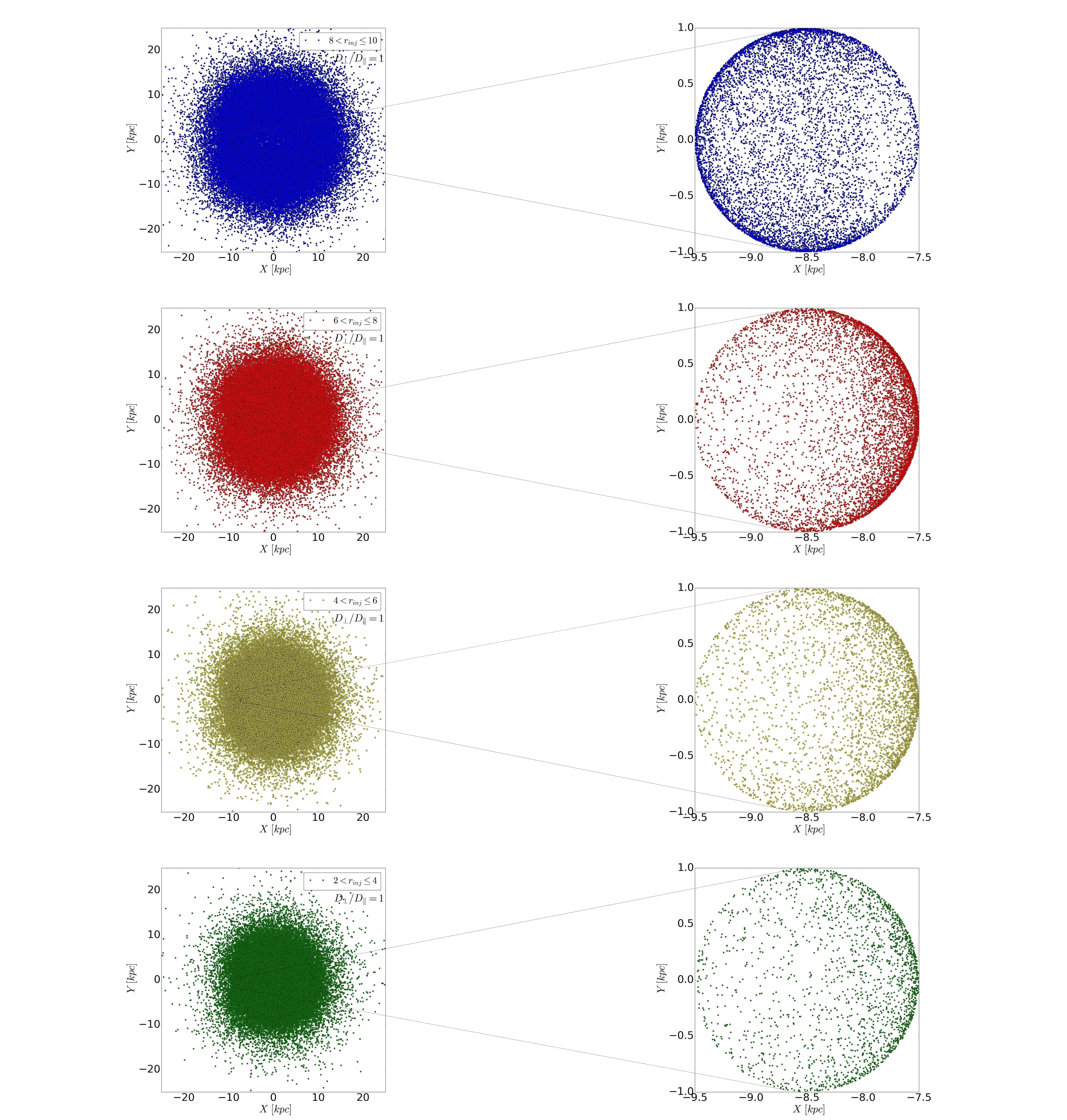}
\caption{This figure shows the position of diffusing CRs in the plane of the disk at the moment they reach the upper or lower boundary at $ z=\pm H_{\rm cr}= 4\;{\rm kpc}$  or $r = r_{\max} = 20\; \rm kpc$, for each rings as indicated in each panel. 
All particles are injected uniformly and randomly in each ring (left column). The right column shows the projected density of CRs inside the local volume around the Solar System for each of the rings. The asymmetry in the distribution reflects the position of the Sun with respect to the rings and shows the general radial gradient in the CR density. 
Here we assume isotropic diffusion ($\Dperp/\Dpar = 1 $) so that there is no
influence of the large-scale (regular) Galactic magnetic field.}
\label{figure1}
\end{figure}

To  investigate the effects of the strength of perpendicular diffusion on the CR flux at Earth from multiple sources we have simulated the propagation 
of CRs for sources that reside in four rings, with a cylindrical radius $r_{\rm inj}$  (in kpc) between
$2 <r_{\rm inj} < 4$, $4 <r_{\rm inj} < 6$, $6 < r_{\rm inj} < 8$ and $8 < r_{\rm inj} < 10$. 
In each case we use two values for the ratio of perpendicular and parallel diffusion coefficients: $\epsilon = \Dperp / \Dpar = 0.01$, 
the case of weak diffusion across the magnetic field where the diffusive step perpendicular to the field ($ \propto \sqrt{D_{\perp}}$)
is ten times smaller than the step along the field,
and $\epsilon =1$, the case of isotropic diffusion where the magnetic field has no influence.
The parallel diffusion coefficient is kept constant at $\Dpar = 3 \times 10^{28}$ cm$^{2}$ s$^{-1}$. The thickness of the CR halo above the disk is
taken to be $H_{\rm cr} = 4 \; {\rm kpc}$. CRs escape freely once they reach $z = \pm H_{\rm cr}$.
 \nskip 
Figure \ref{figure1} (left column) shows the spatial distribution of the CRs in the Galactic disk at the moment of escape, projected onto the disk mid-plane. 
CRs are injected randomly in the four rings with a uniform distribution per unit disk area.
The CRs are recorded once they reach upper (lower) boundary  of the CR halo above the Galactic disk, located at $z\;=+ H_{\rm cr}$ ($z=- H_{\rm cr} $), or 
when they reach the edge of the Galaxy, taken to be at a Galacto-centric distance equal to $r = r_{\max} = 20 \; {\rm kpc}$.

\begin{figure}
\includegraphics [width=\columnwidth]{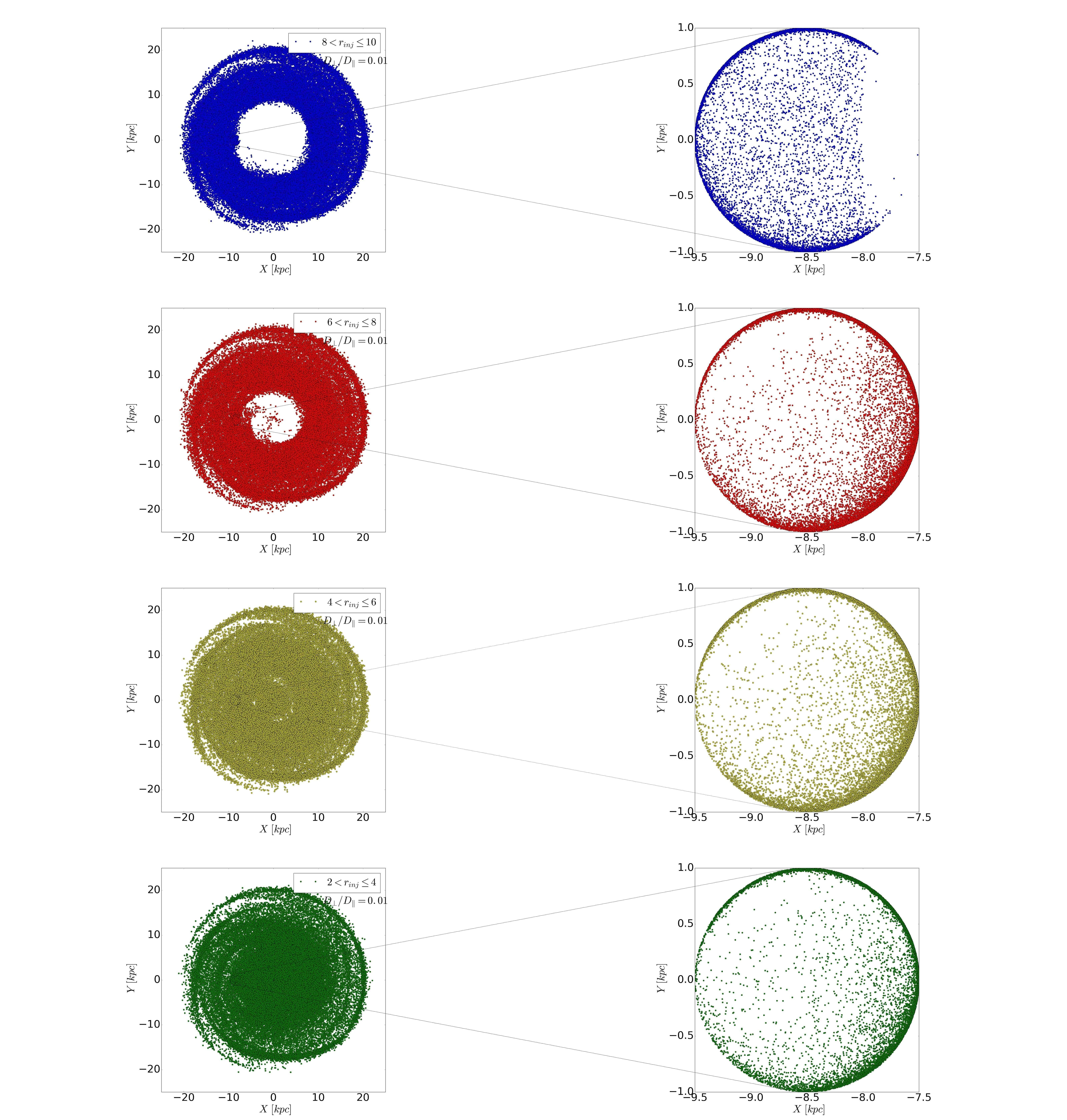}
\caption{As in figure \ref{figure1}, but now for strongly anisotropic diffusion with the value of the ratio $\Dperp/\Dpar=0.01$. One now sees the spiral structure
of the disk field in the CR distribution, as CRs tend to follow magnetic field lines.}
\label{figure2}
\end{figure}

The right column of Figure \ref{figure1} shows, for each of the rings, the disk-projected density of all CRs recorded 
inside the local volume around the Solar System
during the entire simulation. We neglect the height of the Sun ($z_{\odot} \simeq 20 \; {\rm pc}$) above the plane.
The asymmetry in this distribution reflects the position of the Sun with respect to the rings. 
\nskip
Figure \ref{figure2} is the same as Figure \ref{figure1}, but now for strongly anisotropic diffusion with $\epsilon = \Dperp/\Dpar=0.01$.  
As one can see in the right column, top to down: the number of CRs inside the local volume decreases when the sources (on average)
are located farther away from the Sun. This effect becomes more pronounced for small $r_{\rm inj} < r_{\rm \odot} = 8.5 \; {\rm kpc}$. Here we see the influence of
the fact that only those parts of the ring contribute to the CR density that are magnetically connected to the local volume by the spiral disk field in the JF model, since
diffusion across the field is much slower than diffusion along the field in this case.

\begin{figure}
\includegraphics [width=\columnwidth]{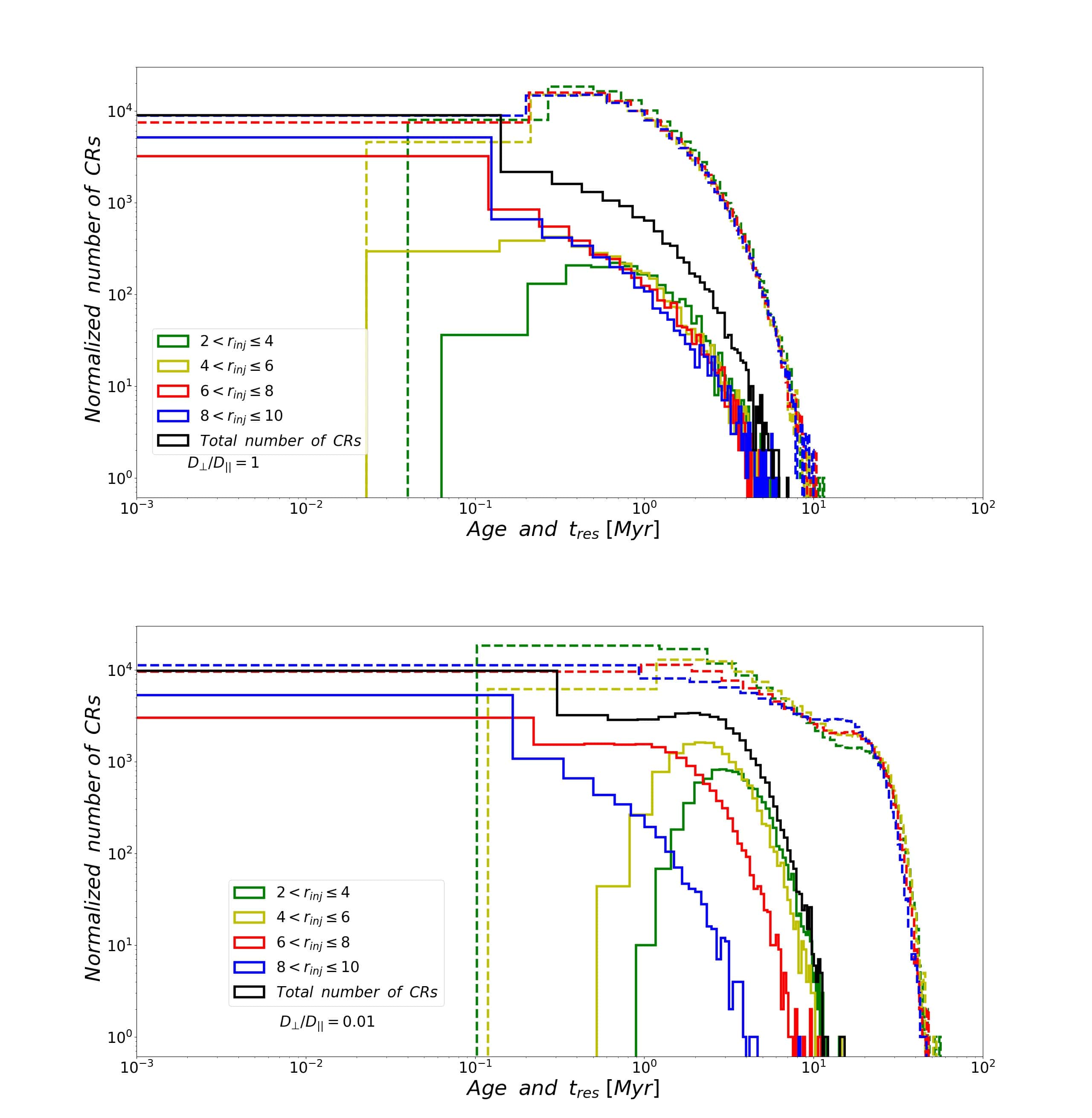}
\caption{This figure shows, for each ring,  the normalized  distribution over age of the CRs observed in the local volume around the Solar System (solid lines), as well as the normalized distribution of the total residence time in the Galaxy (dashed lines). The black solid line shows the total number of CRs. All particles are injected randomly in the four rings with a uniform distribution for the value of the ratio $\Dperp/\Dpar = 1$ (upper panel: isotropic diffusion) and $\Dperp/\Dpar = 0.01$ (lower panel: anisotropic diffusion).}
\label{figure3}
\end{figure}

In both cases we show the results for a {\em fixed} number of simulated CRs (in this case $10^5$ in each simulation) from each ring. If SNRs are distributed
uniformly per unit disk area, the absolute number of CRs produced in each ring is proportional to the ring area, $\pi \left(r_{\rm max}^2 - r_{\rm min}^2 \right)$, if
$r_{\rm min} < r_{\rm inj} < r_{\rm max}$. This determines the relative contributions from the different rings to the total number of CRs at a given location. 
That will be considered below in Figure \ref{figure4}. Below we will also relax the assumption that the SNRs producing the CRs 
are distributed uniformly over the Galactic disk.  
\nskip
In Figure \ref{figure3}  we compare the  distribution of total residence time in the Galaxy (the dashed lines) to the distribution of CR ages
inside  the local volume (the solid lines). We do this separately for CRs coming from each of the four rings. 
The black distribution shows the total number of CRs from all four rings together.

What is shown here (and in the following figures) is the distribution over CR age that is obtained when one sums over the entire duration of the simulation. 
This essentially corresponds to an integration over time of the CR number inside the local volume. 
This is equivalent to the mean distribution that one would observe (at a given time)
for a continuous and constant CR production rate in the Galaxy. 
The simulation runs long enough for all CRs created at the start of the
simulation to have escaped the Galaxy. As before we take $H_{\rm cr} =  4 \; {\rm kpc}$ for the thickness of the CR halo 
and $\epsilon = \Dperp/\Dpar=1 (upper\; plot),\; 0.01 (lower\; plot)$.
\nskip
The distributions shown  in Figure  \ref{figure3} are {\em normalized} distributions, that is: 
they do not take account of the different absolute number of CRs produced in each individual ring.
This is perhaps most obvious in the residence time distributions (the dashed histograms). 
These are virtually identical for each of the four rings. That result is to be expected since [1] the
height $H_{\rm cr}$ of the escape boundary above the Galactic mid plane is taken to be independent of Galacto-centric radius $r$, [2] the diffusion coefficients
are assumed to be constant and [3] the disk component of the spiral disk field in the JF magnetic field model has a constant pitch, independent of radius.
\nskip
In the case of isotropic diffusion, the residence time distribution and the distribution of CR age are very similar in shape, in agreement with the notion that
CRs cross the Galactic disk many many times before escaping. Of course the age of CRs inside the local volume is always less than the total residence time.
The contribution of the ring $8 < r_{\rm inj} < 10 \; {\rm kpc}$, which contains the Solar System,
and the adjacent ring $6 < r_{\rm inj} < 8 \; {\rm kpc}$
show a significant increase of CRs with an age below $10 \;  \rm Myr$  in the local volume: these are CRs injected close to the Sun 
that are able to reach the local volume in a relatively short time. For the rings with $r_{\rm inj} < 6 \; {\rm kpc}$ this effect is not present.

In the case of strongly anisotropic diffusion with $\epsilon = 0.01$, the age distribution (the solid lines) 
is very flat until about $ 10^7\; \rm yr$, for larger ages  the number of particles decreases rapidly. 
In contrast: for $\epsilon = 1$ the number of CRs decreases continuously with increasing age.  
The typical CR age is now also longer than in the case $\epsilon = 1$,
as for most of the disk it is regulated mostly by $\Dperp$ rather than $\Dpar$ so that
$t_{\rm res} \simeq H_{\rm cr}^2/2 D_{\perp}$, see also \citet{ALZetoun2018}. 

This result can be quantified for CRs originating close to the Sun, at $r = R_{\odot} = 8.5$ kpc. 
Escape is mainly through diffusion in the vertical ($z$-direction), with an effective diffusion coefficient equal to:
\be
	D_{zz} = \hatz \bdot \bm{\mathrm{D}} \bdot \hatz = \Dpar \: \left[ \: b_{z}^2 + \epsilon \: \left(1 - b_{z}^2 \right) \: \right] \; .
\ee
Here $b_{z} = B_{z}/B$ is the component of the unit vector along the magnetic field in the $z$-direction.
Using the numbers in \citet{JF2012A}: near the Sun one has a spiral disk field with strength equal to $B_{\rm disk} = 1.18 \; \mu{\rm G}$ and pitch angle
$p = 11.5$ degrees, an {\em X-field} in the $r-z$ plane with a strength
of $B^{\rm X} \simeq 0.25 \: \mu{\rm G}$ and a field elevation angle of $i_{0} = 49$ degrees. 
This implies $B_{r} = B_{\rm disk} \: \sin p + B^{X} \: \cos i_{0} \simeq 0.40 \; \mu{\rm G}$, $B_{\phi} = B_{\rm disk} \: \cos p \simeq 1.16 \; \mu{\rm G}$ and
$B_{z} = B^{\rm X} \: \sin i_{0} \simeq 0.19 \: \mu{\rm G}$. Then $B \simeq 1.24 \; \mu{\rm G}$ and $b_{z} = B_{z}/B \simeq 0.11$ so that:
\be
	D_{\rm zz} = \Dpar \: \left[ \: 0.023 +   0.976 \: \epsilon \right] \simeq 0.033 \; \Dpar \; ,
\ee
where the last equality is for $\epsilon = 0.01$. In that case the effective residence time increases with respect to the case of isotropic diffusion to 
$t_{\rm res} = H_{\rm cr}^2/2D_{zz} \simeq 30 \: H_{\rm cr}^2/2 \Dpar$, in
good accordance with the results of the simulations shown in Figure \ref{figure3}.

\nskip
The distribution of CR age of CRs inside the local volume no longer resembles the residence time distribution. 
For the two inner rings, $2 < r_{\rm inj} < 4 \; {\rm kpc}$ and
$4 < r_{\rm inj} < 6 \; {\rm kpc}$ (the green and yellow histograms respectively), CRs take much longer to reach the local volume as they have to diffuse a considerable
distance across the small-pitch JF disk magnetic field spiral. This explains the absence of particles from these rings at the lower end of the age distribution.
CRs originating from the ring with $6 \; {\rm kpc} < r < 8 \: {\rm kpc}$ (red histogram) have an age distribution that is very flat until $(2-3)\; \rm Myr$, 
decaying rapidly thereafter. CRs injected from $8 \: {\rm kpc} < r < 10 \: {\rm kpc}$ show a strong decrease with increasing age, 
which is the combined effect of the
pitch of the disk field and of the shape of the X-field which transports particles to larger Galacto-centric radii through parallel diffusion along the resultant magnetic field.
In all cases the CRs in the local volume are now considerably younger than escaping particles, much more so than in the case of isotropic diffusion.

Summarizing: both the residence time distribution and the age distribution of CRs in the local volume depend rather strongly 
on the  diffusion coefficient ratio $\epsilon = \Dperp/\Dpar$. 
When $\Dperp / \Dpar = 1$ the particles diffuse  relatively fast, and  need less time to reach  
Earth or escape the Galaxy. In this case, the two distributions have a similar shape. When $\Dperp / \Dpar = 0.01$ CRs reside $\sim 50 \times$ longer in the Galaxy.
The escape time distribution and the age distribution in the local volume no longer resemble each other, and the typical age of CRs in the local volume
is considerably less than the residence time. CRs originating from $r_{\rm inj} \le 6 \; {\rm kpc}$ are only able to reach the local volume after
(typically) $10 \; {\rm Myr}$.
\nskip
Figure \ref{figure4} shows the absolute number of simulated CRs from each of the four rings 
found over the entire simulation inside the local volume  as a function of CR age. 
Here we assume a
source density per unit disk area so that the number of CRs produced in each ring is proportional to the ring area $\pi (r_{\rm max}^2 - r_{\rm min}^2)$.
The black distribution is the  total number of CRs from all four rings, which is what determines the local CR flux. The parameters of the simulation are the same as before.

In the case of isotropic diffusion ($\epsilon = 1$) one clearly sees that the inner rings contribute less to the local CR density at given CR age. 
In the anisotropic case
($\epsilon = 0.01$) the contributions from each of the rings are very distinct due to the effects of the magnetic field geometry on CR propagation.
\nskip
Figures \ref{figure5}  and \ref{figure6} are similar to Figure \ref{figure4}. Here  we  assume that the surface density $N(r)$ of 
CR sources (SNRs) scales with Galacto-centric radius $r$ as: 
\be
\label{W1}
	N_{\rm snr} (r) \propto  {\rm exp}\left( - \frac{r}{R_{\rm snr}} \right) \; .
\ee
The exponential scale length $R_{\rm snr}$ is chosen to be $R_{\rm snr} \;= 5.4 \;{\rm kpc}$ in Figure \ref{figure5} and 
equal to $R_{\rm snr} = 8 \; {\rm kpc}$ in Figure \ref{figure6}. 
All simulated CRs are given a weight $\propto N(r_{\rm inj})$, with $r_{\rm inj}$ the injection radius.
The solid-line histograms (in color) give the number of CRs in the local volume from each of the rings. The total number 
of CRs from all rings is shown in black, all as a function of CR age.

\nskip
Figures \ref{figure7}/\ref{figure8}  are similar to Figures \ref{figure5}/\ref{figure6}, but for the SNR 
surface density defined by \citet{Green2015} and \citet{Sasaki2004}:
\be
\label{W2}
	N_{\rm snr}(r) \propto  \left(\frac{r}{{R}_{\odot}} \right)^{\alpha} \;   {\rm exp}\left( - \frac{r}{R_{\rm snr}} \right) \; .
\ee
Here $\alpha = 1.1$, ${R}_{\odot} =8.5 \; {\rm kpc}$ is the Galacto-centric radius of the Solar orbit around the Galactic Center 
and we take $R_{\rm snr} \;= 5.4 \;{\rm kpc}$ in Figure \ref{figure7} and $R_{\rm snr} \;= 8  \;{\rm kpc}$ in
Figure \ref{figure8}.

\nskip
Figure  \ref{figure9} shows how the CRs inside the local volume around the Solar System are  
distributed over the accumulated  grammage.  Again we integrate over the entire simulation run. 
The contribution from each of the  rings (solid colored lines) is shown separately, as well as its sum (solid black line). The
grammage distribution at the moment of escape for the Galaxy is shown in dashed lines. For the grammage calculation we assume 
the smooth density distribution of Eqn. (\ref{grammage}). Simulation parameters are the same as before.
The accumulated grammage becomes larger when the source is farther from the Sun. 
It also increases when the value of the ratio $\epsilon = \Dperp/\Dpar$ is small, 
because the CRs take longer to escape from the Galaxy, as discussed in some detail above.

\subsection{Results for CRs from four hypothetical sources at the same distance from  Earth }
\label{sec:3.2}

In order to investigate the effects of the strength of perpendicular diffusion on CR numbers at Earth 
we have simulated the propagation of CRs from four discrete equidistant sources (source distance: $5 \; \rm kpc$ from Earth).
These sources are typically located on different sections of the JF spiral disk field, see Figure \ref{figure10}, bottom panel.
It is assumed that the sources inject the CRs simultaneously at $t = 0$.
Again we perform simulations for two values of the ratio $\epsilon = \Dperp / \Dpar = 0.01,\;1$. The value of the
parallel diffusion coefficient is kept constant at $\Dpar = 3 \times 10^{28}$ cm$^{2}$ s$^{-1}$. The thickness of the CR halo above the disk is taken to be $H_{\rm cr} = 4 \; {\rm kpc}$. CRs escape freely once they reach $z = \pm H_{\rm cr}$.
Corresponding values for CR ages in the local volume
for other values of $D_{\parallel}$ can be obtained from the scaling law $t \propto D_{\parallel}^{-1}$.

\nskip
The top two panels in Figure \ref{figure10} show, as a function of time, the {\em normalized} number of CRs inside the local volume 
for these four  hypothetical  sources. The black distribution shows the total number of CRs from all four sources. 
When $\epsilon = 1$  (the case of isotropic diffusion) the distribution of the CRs over time from all four sources is identical, 
apart from statistical fluctuations
due to the finite number of simulated CRs per source. Therefore, all  four sources
contribute equally to the local CR density at Earth, as expected for this case.
The CRs also escape relatively quickly from the Galaxy.

In contrast: for  $\epsilon = 0.01$ the distribution over time becomes wider as the CRs take more time 
to escape from the Galaxy, as already discussed.
By far the largest contribution now comes from Source 3 that lies on the same section of the JF spiral disk field as the Sun. 
CRs from that source can reach
Earth relatively rapidly through parallel diffusion.
The contribution from Sources 1 and 2 is reduced as a significant amount of cross-field diffusion is required, taking a longer time. 
There is no contribution from Source 4: CRs from this source have not enough time to diffuse the required distance across the magnetic field 
to reach the local volume before escaping from the Galaxy.

\subsection{Results for CRs from ten specific sources close to Earth }
\label{sec:3.3}

Next we investigate the effect of the strength of perpendicular diffusion on the CR flux at Earth from  ten specific nearby discrete sources.
All sources lie within a distance of $1.5\; {\rm kpc}$. 
We employ the same source set as the one employed by \citet{Thoudam2007}. Simulation parameters are the same as before.
 
\begin{table}
\centering
\begin{tabular}{|c|c|c|c|c|}
\hline
\textbf{ SNRs} & \textbf{ Distance [kpc]} & \textbf{Galactic longitude ($l$) [degree]} &\textbf{Age[kyr]}& \textbf{Position(X,Y,Z)[ kpc]} \\
\hline
\hline
CTA1 & 1.40 &119.5&24.5& (-9.18, 1.20, 0.28) \\
\hline
G65.3+5.7 & 1.00& 65.3 & 14&(-8.17, 0.72, 0.10)\\
\hline
G73.9+0. 9& 1.30 & 73.9 & 10&(-8.14, 1.25, 0.05)\\
\hline 
HB21 & 0.80 &89.0 &19& (-8.48, 0.80, 0.10) \\
\hline
G114.3+0.3 & 0.70& 114.3 &41& (-8.79, 0.64, 0.03)  \\
\hline
HB9 & 1.00 &160.9&7.7& (-9.44, 0.33, 0.08)\\
\hline
S147 &  0.80 &180.0& 4.6&(-9.3, 0.00, 0.00) \\
\hline
Vela & 0.30 & 263.9 &11&(-8.53, -0.30, 0.01)  \\
\hline
G299.2-2.9 & 0.50&299.2 &5&(-8.26, -0.44, 0.00 )\\
\hline
Cygnus Loop & 0.44 &74.0&14& (-8.38, 0.42, -0.03 )  \\
\hline
\hline

\end{tabular}
\caption{Parameters of  known SNRs located within a distance of 1.5 kpc from the Solar System considered in our calculation. }
\label{table:1}
\end{table}
\nskip 
For  such nearby discrete sources, one expects significant temporal fluctuations in the local density of CRs at Earth. 
The exact position of these SNRs and the direction of the local  Galactic magnetic field are shown in Figure  \ref{figure11}. 
In order to simplify things we assume that each source
produces the CRs impulsively at the time of the supernova explosion. 
In reality, most of the CRs are produced in a few Sedov-Taylor times $ t_{\rm ST}$, given for typical SNR parameters by:
\be
	t_{\rm ST} \simeq \frac{R_{\rm dec}}{V_{\rm f}} = \frac{\displaystyle \left( 3 M_{\rm ej}/4 \pi \rho_{\rm ism} \right)^{1/3}}
	{\displaystyle \sqrt{2 E_{\rm snr}/M_{\rm ej}}}  \simeq 1.7 \times 10^{3} \: \left( \frac{E_{\rm snr}}{10^{51} \; {\rm erg}} \right)^{-1/2}
	\left( \frac{n_{\rm ism}}{1 \; {\rm cm}^{-3}} \right)^{-1/3} \: \left( \frac{ M_{\rm ej}}{10 \; \Msun} \right)^{5/6} \; {\rm yr} \; .
\ee

Here $E_{\rm snr}$ is the mechanical energy driving the expansion of the supernova remnant, $M_{\rm ej}$ is the ejecta mass, 
$V_{\rm f} = \sqrt{2 E_{\rm snr}/M_{\rm ej}}$ is the free-expansion velocity of the remnant for $t < t_{\rm ST}$,
$R_{\rm dec}$ is the deceleration
radius where the SNR has swept up an amount of mass equal to the ejecta mass in interstellar material with number density $n_{\rm ism}$.  
Since this is much shorter than the typical ($10^7 \; yr$) CR residence time
this simplification is justified.
\nskip  
In Figure  \ref{figure12} we show, for each of the sources of Table \ref{table:1}, the distribution of  the normalized number of CRs 
that reside in the local volume around the Solar system as a function of time. Again we consider the two cases: $\Dperp/\Dpar=\; 1$ (upper plot), $\Dperp/\Dpar=\; 0.01$ (lower plot). 
The dashed distribution shows the total number of CRs from all sources. The associated grammage distribution is shown in Figure \ref{figure13}. The dashed distribution again is the total number of CRs from all ten SNRs sources. Again we choose $\Dperp/\Dpar=\; 1$ (upper plots), $\Dperp/\Dpar=\; 0.01$ (lower plots). Since these sources are relatively close,
the difference between the case $\Dperp/\Dpar=\; 1$ and the case $\Dperp/\Dpar=\; 0.01$ is much less pronounced than in the other simulations presented here.
Their location (shown in Figure \ref{figure11}) is such that for most of the sources the CRs can diffuse mostly along the local magnetic field to reach the Local Volume.
The exceptions are S147 and HB9. In the case $\Dperp/\Dpar=\; 0.01$ they show the tell-tale sign of an absence of \lq{}young\rq{} CRs in the lower panel of Figure \ref{figure12}, indicating that the CRs from these sources take longer to reach Earth when diffusion across the field is slow.

\subsection{Model test: the Boron to Carbon ratio}
\label{sec:3.4}

The Boron to Carbon ratio ${\rm B/C}$ is widely used to determine key parameters (diffusion coefficient(s), scaling with CR energy etc.) of propagation models by comparing model predictions with observations.  The Boron to Carbon ratio is (of many possible ratios of secondary to primary cosmic rays nuclei) the easiest to measure experimentally.
Recently, this ratio has been determined with an accuracy of a few percent by the 
Alpha Magnetic Spectrometer (AMS-02) on the International Space Station \footnote{http://www.ams02.org}, 
over a relatively wide range of energies. Several other experiments like PAMELA \footnote{https://www.ssdc.asi.it/pamela/},
and CREAM  \footnote{https://cosmicray.umd.edu/iss-cream/}  have also  measured the B/C ratio, 
in a range of kinetic energies between   $100 \; \rm MeV$ and  $1  \; \rm TeV/nucleon $.
\nskip

We have used the weighted slab method using the Grammage  distributions for CR  protons inside the local volume 
from the four rings to calculate  B/C ratio as a function of kinetic  energy per nucleon, using the path length distribution obtained from the simulations. 
In our calculation we have chosen the value of the parallel diffusion coefficient $D_{\parallel}$ 
so that the B/C ratio obtained coincides with the observed value at  $1 \; \rm GeV/nucleon$. For its scaling with energy ($D_{\parallel} \propto E^{\delta}$)
we use a value of $ \delta = 0.33$, the value for standard Kolmogorov turbulence, which gives a reasonable fit to the observational data.

\nskip
For these simulations this procedure is allowed since we have assumed that $\epsilon = D_{\perp}/D_{\parallel}$ is constant. It should be pointed out that this assumption, although convenient, is often not justified since in many cases $D_{\parallel}$ usually scales differently with CR energy than $D_{\perp}$. 
This was already discussed briefly in Section \ref{sec:2.2} of this paper. A non-trivial scaling of $\epsilon = D_{\perp}/D_{\parallel}$ with CR energy 
introduces yet another parameter in the CR propagation model that has to be determined
from theory and/or observations.

\nskip
With the weighted slab method we calculate the Boron to Carbon abundance ratio as a function of energy per nucleon. 
As before we consider CRs in the local volume  around the Solar System.
In our calculation we assume that only Primary CRs  nuclei  O, N, and  C serve as parent  nuclei for B.  
The source abundance of each of the parent nuclei is taken from \citet{Strong2001}. 
Relevant nuclear data, such as cross sections, are found in \citet{Webber2003}, \citet{Garcia-Munoz1987}, and  \citet{Ramaty1997}.
As before we consider primary CRs that originate from the four rings centered on the Galactic Center.
Figure \ref{figure14}  shows the  B/C abundance ratio, plotted as a function of energy per nucleon
for two cases:  isotropic diffusion with $D_{\perp} = D_{\parallel}$ (Figure 14a) and strongly anisotropic diffusion with $D_{\perp} = 0.01 \: D_{\parallel}$ (Figure 14b).
The observed ratio is shown for comparison.
Figures 14c and 14d  show the {\em total} number of CRs in the local volume, adding the contributions
 from the four rings, for $\epsilon = 1$ and for $\epsilon =  0.01$,  as indicated in each panel.  One clearly sees a good agreement  with observations. 
This in itself is not too surprising: in this cases the path-length distribution as a function of energy is set by the (identical) scaling of
$D_{\parallel}$ and $D_{\perp}$ with CR energy per nucleon. By choosing $D_{\parallel}$ at $1\; \rm GeV/nucleon$  in such a way that the calculated Boron to Carbon ratio matches the
observed value hides the effect of a varying anisotropy in CR diffusion.
We have adjusted the value of $ D_{0}$ 
to produce the best fit with the observational data. We use
$ D_{0}= 3 \times 10^{28}$ cm$^{2}$ s$^{-1}$ for strongly anisotropic diffusion with $\epsilon = \Dperp/\Dpar=0.01$, and $D_{0}= 2.8 \times 10^{27}$ cm$^{2}$ s$^{-1}$ for isotropic diffusion with $\epsilon = \Dperp/\Dpar=1$. This reduction by an order of magnitude of $D_{0}$ in the latter case reflects the need to keep the
typical residence time the same in order to fit the observations, see the related discussion in Section \ref{sec:3.1}.

\nskip
To illustrate this point: in Figure \ref{figure15}  we show the effect of varying the ratio $\epsilon = D_{\perp}/D_{\parallel}$, but now (unlike Figure \ref{figure14}) for 
a {\em constant} value of  $\Dpar =4 \times 10^{28}$ cm$^{2}$ s$^{-1}$. As expected the B/C ratio increases with  decreasing $\epsilon$: when $\epsilon$ is small (primary) CRs reside longer inside the Galaxy, 
thereby accumulating a larger grammage and producing more secondary CRs.
\nskip
In our calculation we did not address the calculation of the B/C ratio in the energy below $1\; \rm GeV/nucleon$.  It was found,  see for instance \citet{Garcia-Munoz1987}, and \citet{Webber2016}, that the ratio
increases with energy below $1\; \rm GeV/nucleon$, and decreases above this energy, instead of the monotonic decrease with energy expected in the simplest Galactic diffusion model.
Below $1\; \rm GeV/nucleon$, the B/C ratio may be influenced  to a significant  degree  by the advection of CRs by a Galactic wind, see for instance \citet{Strong1998}, or by CR re-acceleration during the propagation through the Galaxy,
 see for instance \citet{Simon1986}. Alternatively, this behavior can be explained by having a constant CR diffusion coefficient below a few $ \rm GeV/nucleon$.
 
\subsection{Model test: fit to the CR data}
\label{sec:3.5}

The spectrum of CR nuclei  appeared to be consistent with a single power law in particle rigidity as defined by \citet{Longair2011} $R_{\rm cr} = pc/Ze$, $J_{\rm cr} \propto R^{-\delta}_{\rm cr}$  with slope $\delta \approx  2.7$ in the energy range from $\sim \rm  GeV$ to few $\rm PeV$.
 
\nskip
Recent experiments  of (AMS-02) on the International Space Station,  and  PAMELA  have measured the energy spectra of   various elements of CRs  from protons to heavier nuclei  below $10^{6}\; \rm GeV$.

\nskip
In Figure \ref{figure16} we show  the spectrum of Proton, Helium and Carbon nuclei $ \times E^{2.7}$  as function of kinetic energy per nucleon for the  case of the diffusion coefficient $\Dperp/\Dpar= 0.01$ compared   by recent experiments  of 
(AMS-02) on the International Space Station,  and  PAMELA, for proton: \citet{pAMS2015}, \citet{pPAMELA2013}, for He: \citet{CHe2017}, \citet{HePAM2011}, for C: \citet{CHe2017}, \citet{CPam2014}.
The best fit to the spectrum according to a power law is represented by the color solid line as indicated in the plot.  One clearly sees a good agreement  with observations. 

\nskip
Secondary CR nuclei are created by spallation of primary CRs that interact with matter during their journey through the Galaxy by means of nuclear interactions of heavier fragments with the ISM. Spallation processes will remove nuclei according to:
\be
\label{survival}
\frac{{\rm d} N}{{\rm d} \ell} = -\sigma_{\rm sp} \; n_{\rm H}( \bm{r} )\; N.
\ee

where ${\rm d} \ell = v\; {\rm d}t$ is the path length increase in a time-span ${\rm d}t$, $n_{\rm H}(\bm{r})$ is the Hydrogen number
density at CR position $\bm{r}$ 
and $\sigma_{\rm sp}$ is the total spallation cross section.
The formal solution of this equation is:
\be
\label{survival2}
	N(\ell)= N(0) \;  {\rm e}^{-\lambda(\ell)} \; \; , \; \; \lambda(\ell) \equiv \frac{ \sigma_{\rm sp} \;   \Sigma_{\rm cr}(\ell) }{\mu \; m_{\rm H}} \; .
\ee
The total  spallation cross sections are found, for instance in \citet{Longair2011}.  $\Sigma_{\rm cr}(\ell)$ is the accumulated grammage, 
$\mu  \simeq 1.3$ is the mean mass per particle in units of the Hydrogen mass of the interstellar gas, and $m_{\rm H}$ is the the mass of the Hydrogen atom. 

\nskip
The survival probability is defined (from Eqn. \ref{survival2}) as: 
\be
P(\ell) \equiv \; {\rm e}^{-\lambda(\ell)}.
\ee
In Figure \ref{figure17} we show the survival probability $P(\ell)$ for five elements (Helium, Carbon, Oxygen, Silicon and Iron) 
as a function of grammage. 
The survival probability decreases for heavier nuclei due to an increase in spallation cross section with atomic mass number $A$: 
roughly $\sigma_{\rm sp} \propto A^{2/3}$.


\section{Conclusions}
\label{sec:4}

In the model presented in this work we have studied the effects of anisotropic diffusion on the propagation
and residence time of GCRs. We also considered the
properties of the GCRs in a local volume around the Solar System.
We considered  CRs from sources (supernova remnants) with a smooth distribution of sources, located in different intervals in the Galacto-centric radius.
We also considered the contribution from  distinct sources, located at a fixed distance from the Sun, and from a set of specific, real-world sources  
within a distance of $1.5\; {\rm kpc}$ from the Solar System. All simulations employed the same approach, based on the solution of  stochastic differential equations, 
to describe the CR propagation in the Galaxy. This study of anisotropic diffusion has used the field model of Jansson and Farrar for the
regular (large-scale) Galactic magnetic field.

\nskip
Our simulations show that for a smooth distribution of CR sources (taken to be Type II supernova remnants) the distribution of CR age
inside the local volume around the Solar System, and the residence time distribution at the moment of CR escape from the Galaxy, are fairly similar in shape 
when diffusion is isotropic ($\Dperp = \Dpar$).  They become rather different when diffusion is strongly anisotropic ($\Dperp = 0.01 \: \Dpar$).  Because of the
tightly-wound spiral geometry of the Jansson-Farrar field in the plane of the Galactic Disk CRs originating 
from Galacto-centric distances larger than the Solar distance ($\sim 8.5 \; {\rm kpc}$)
are (relatively speaking) depressed in the case of anisotropic diffusion. They must diffuse a significant distance across the large-scale magnetic field, 
which takes longer for small $D_{\perp}$, allowing a larger fraction to escape from the Galaxy before reaching the local volume.
Given the field geometry this effect is not completely compensated by the larger residence time for the CRs in this case.

\nskip
In addition we considered CRs that originate from ten discrete sources with known age and position  inside the local volume, listed in Table 1. 
For this case we find the following: 
for somewhat distant sources (distance: $5\; \rm kpc$, comparable to the assumed thickness $H_{\rm cr} = 4\; \rm kpc$ of the CR halo)
the effect of anisotropic diffusion on the distribution of CR age and accumulated grammage inside the local volume is still visible, but not as pronounced as
in the case where we consider CRs originating from the whole Galactic disk. For the subset of nearby
 sources (sources with distance less than $1.5\;\rm kpc$) 
the differences between the case of isotropic diffusion and anisotropic diffusion are relatively small. If diffusion is very anisotropic
($\epsilon = 0.01$) sources not connected to the local volume by a field line of the Jansson-Farrar field
near the disk mid-plane field contribute little to the local CR flux.

\nskip
We have shown that for a constant (energy-independent) value of $\epsilon = D_{\perp}/D_{\parallel}$ our code reproduces to observed Boron to Carbon ratio
provided one normalizes $D_{\parallel}$ in such a way that the value of B/C at $1\; \rm GeV/nucleon$ matches the observed value. 
This conclusion will change as soon as one allows $D_{\perp}$ and $D_{\parallel}$ to scale differently with CR energy/nucleon.

\nskip

The results of this paper demonstrate that it is relatively simple to simulate anisotropic CR diffusion in a large-scale magnetic field.
Of course, a more realistic model will have to include additional physical effects that are now commonly included in CR propagation codes. 
In what follows we mention a few of these effects.

\nskip
First of all we have neglected any CR energy changes during propagation.
It is often assumed, see for instance that CRs are re-accelerated after leaving their source,
either by stochastic acceleration in interstellar hydromagnetic turbulence (Fermi-II acceleration) or during random 
encounters with expanding supernova remnants. It has been suggested by \citet{DruStr2017} that as much as 50\% of the power needed to maintain the
Galactic CR population against losses is provided by re-acceleration during propagation through the ISM.
In either case the evolution of the momentum distribution $f(p) = {\rm d}N/(4 \pi p^2 \: {\rm d}p)$ can be described by a diffusion equation in momentum space
of the type:
\be
\label{momdiff}
	\left( \frac{\partial f}{\partial t} \right)_{\rm acc} = \frac{1}{p^2} \: \frac{\partial}{\partial p} \left( p^2 \: D_{p} \: \frac{\partial f}{\partial p} \right) \; .
\ee
Here $D_{p}$ is the momentum diffusion coefficient. Re-acceleration is easily incorporated into codes such as this by adding CR momentum as an additional
dimension (in effect working in phase space) and integrating the following It\^o equation for CR momentum in addition to Eqn. (\ref{TPmotion}):
\be
\label{Itomom}
	\Delta p = \left< \frac{{\rm d} p}{{\rm d}t} \right> \: \Delta t + \sqrt{2 D_{p} \: \Delta t} \: \xi_{p} \; .
\ee
Here
\be
	\left< \frac{{\rm d} p}{{\rm d}t} \right> \equiv  \frac{1}{p^2} \: \frac{\partial}{\partial p} \left( p^2 \: D_{p} \right)
\ee
is the mean momentum gain due to the Fermi-II process and $\xi_{p}$ is another random variable, with zero mean and unit variance. 
The effect of energy losses (ionization losses, expansion losses etc.) are
easily added to the regular momentum gain term, the first term on the right-hand side of Eqn. (\ref{Itomom}).

\nskip
In the second place: we have employed the simplest possible model, where $\Dperp/\Dpar$ is a constant and independent of CR energy. 
The precise form of perpendicular diffusion, determining the scaling of $\Dperp$ with CR energy and its relation to $\Dpar$, 
is a complicated problem with a range of possible regimes and scalings.

\nskip
Finally: this model is (by design) a test-particle model. This makes it difficult to incorporate the effect of scattering of CRs by
{\em self-generated} hydromagnetic waves if wave generation depends on the details of the CR energy distribution or the CR number density.

\nskip
In a follow-up to this research we will consider two of the effects also neglected in these simulations: 
the influence of a Galactic wind and drift of CRs in the magnetic field.

\clearpage

 \clearpage
 \clearpage
 \clearpage
\begin{figure}
\includegraphics [width=\columnwidth]{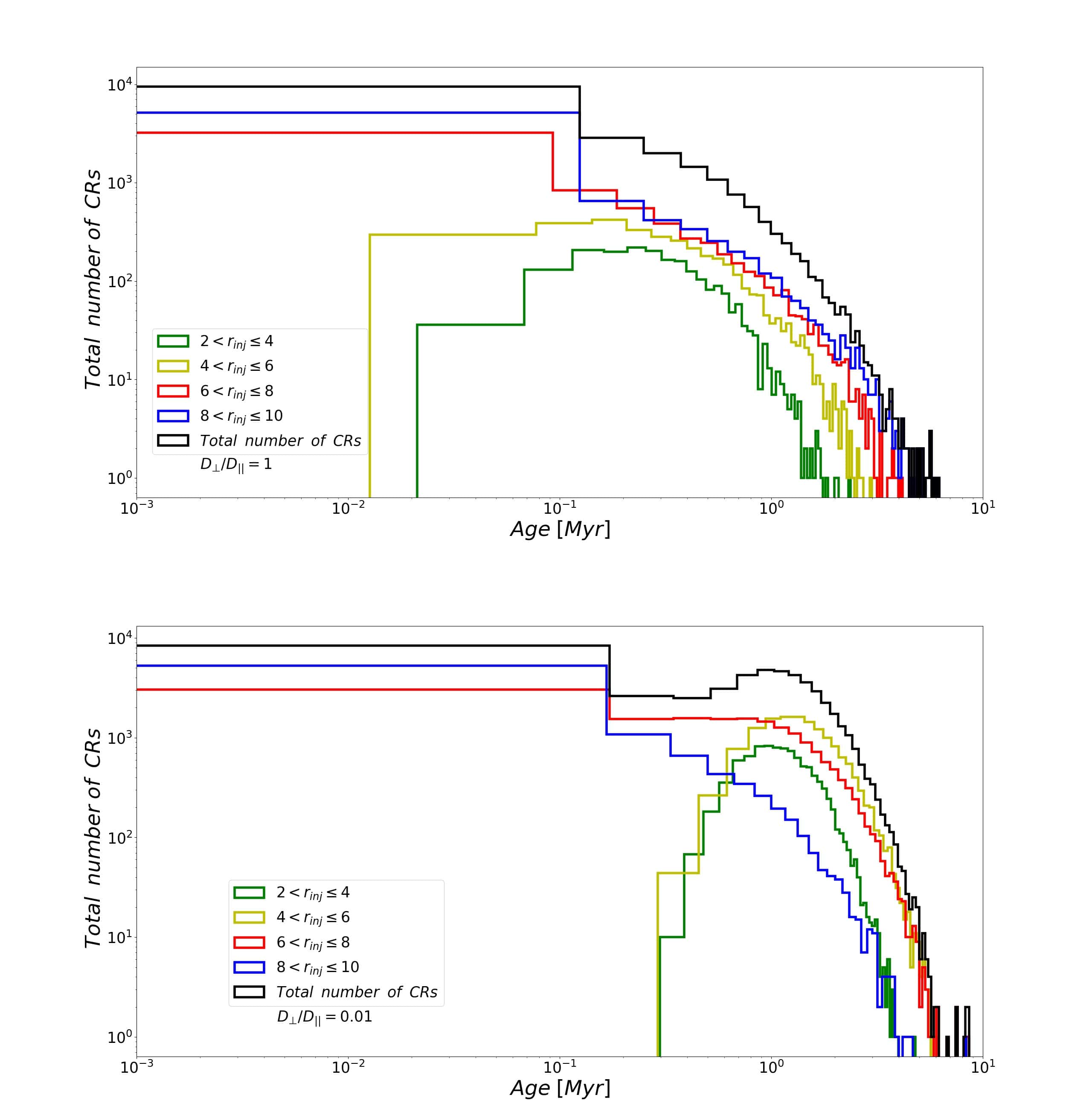}
\caption{This figure shows, for each of four rings, the age distribution of the total number of CRs per unit disk area. 
The black solid line shows the total CRs come from all rings together. 
In each of the four rings CRs are injected randomly with a uniform distribution of sources (Type II supernova remnants) over disk area. Results are shown for $\Dperp/\Dpar=1$ (upper panel: isotropic diffusion) and for $\Dperp/\Dpar = 0.01$ (lower panel: anisotropic diffusion). Note that in the latter case 
for the two inner rings ($r_{\rm inj} \le 6$ kpc) there are no CRs reaching the observer with a small age: the slow diffusion across the field prevents this. }
\label{figure4}
\end{figure} 
 \clearpage
\begin{figure}
\includegraphics [width=\columnwidth]{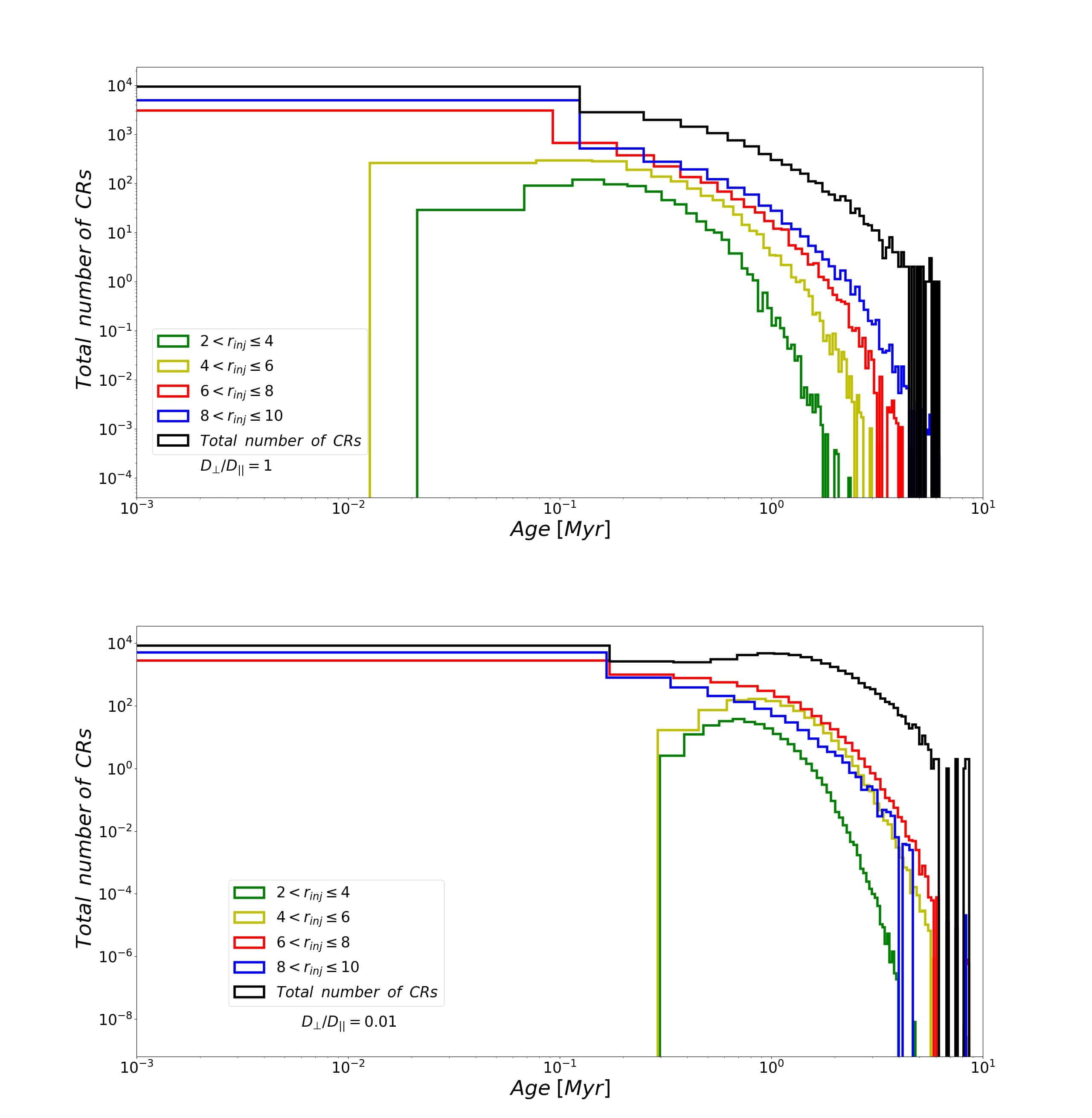}
\caption{ As in figure \ref{figure4}, but using the exponential surface density  distribution (\ref{W1}) for the number of Type II supernova remnants.
In this figure we employ a value of $R_{\rm snr} \;= 5.4 \;{\rm kpc}$. }
\label{figure5}
\end{figure} 
 \clearpage
\begin{figure}
\includegraphics [width=\columnwidth]{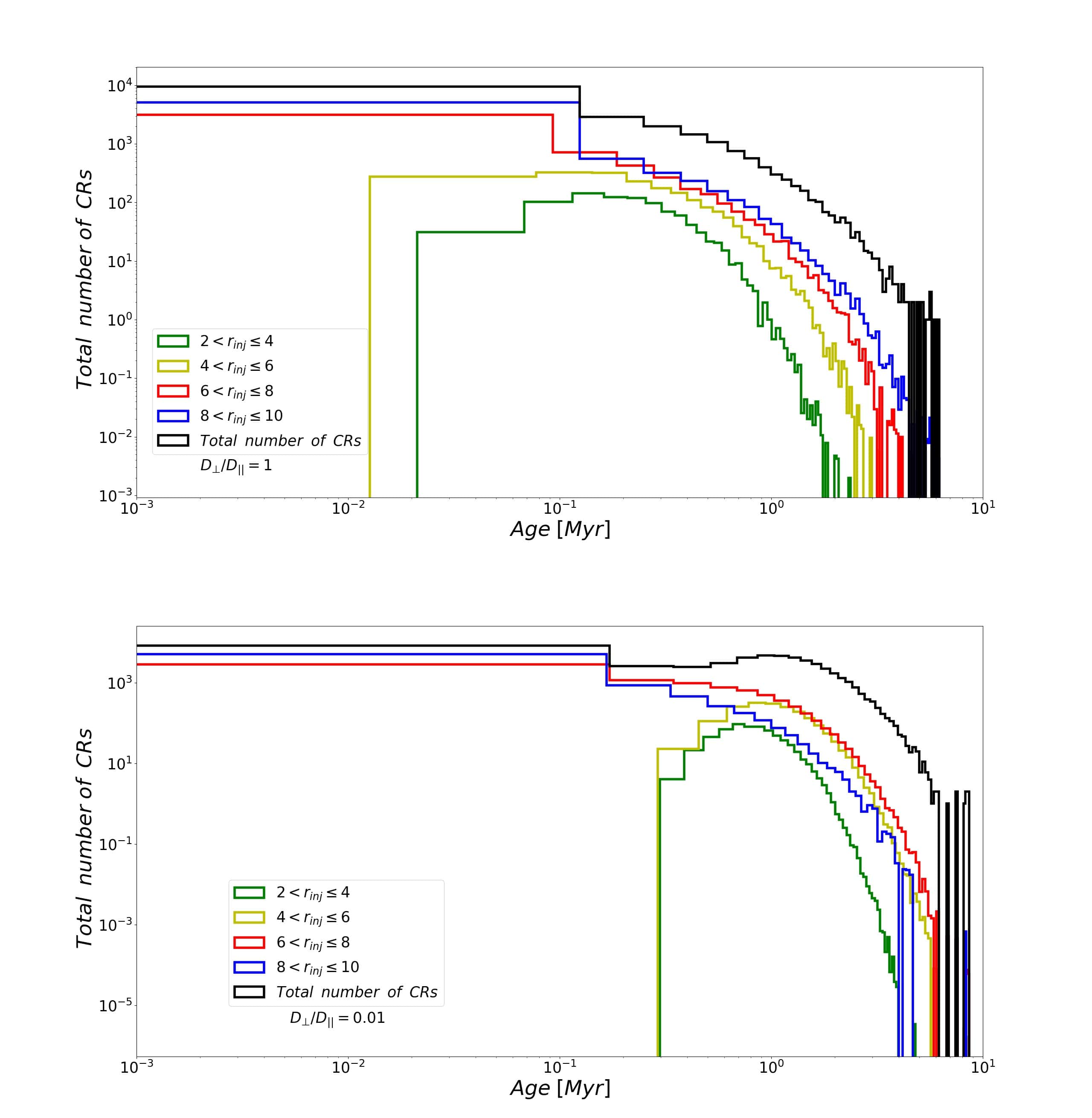}
\caption{As in figure \ref{figure4},  but now for the value of $R_{\rm snr} \;= 8 \;{\rm kpc}$. }
\label{figure6}
\end{figure} 
\clearpage
\begin{figure}
\includegraphics [width=\columnwidth]{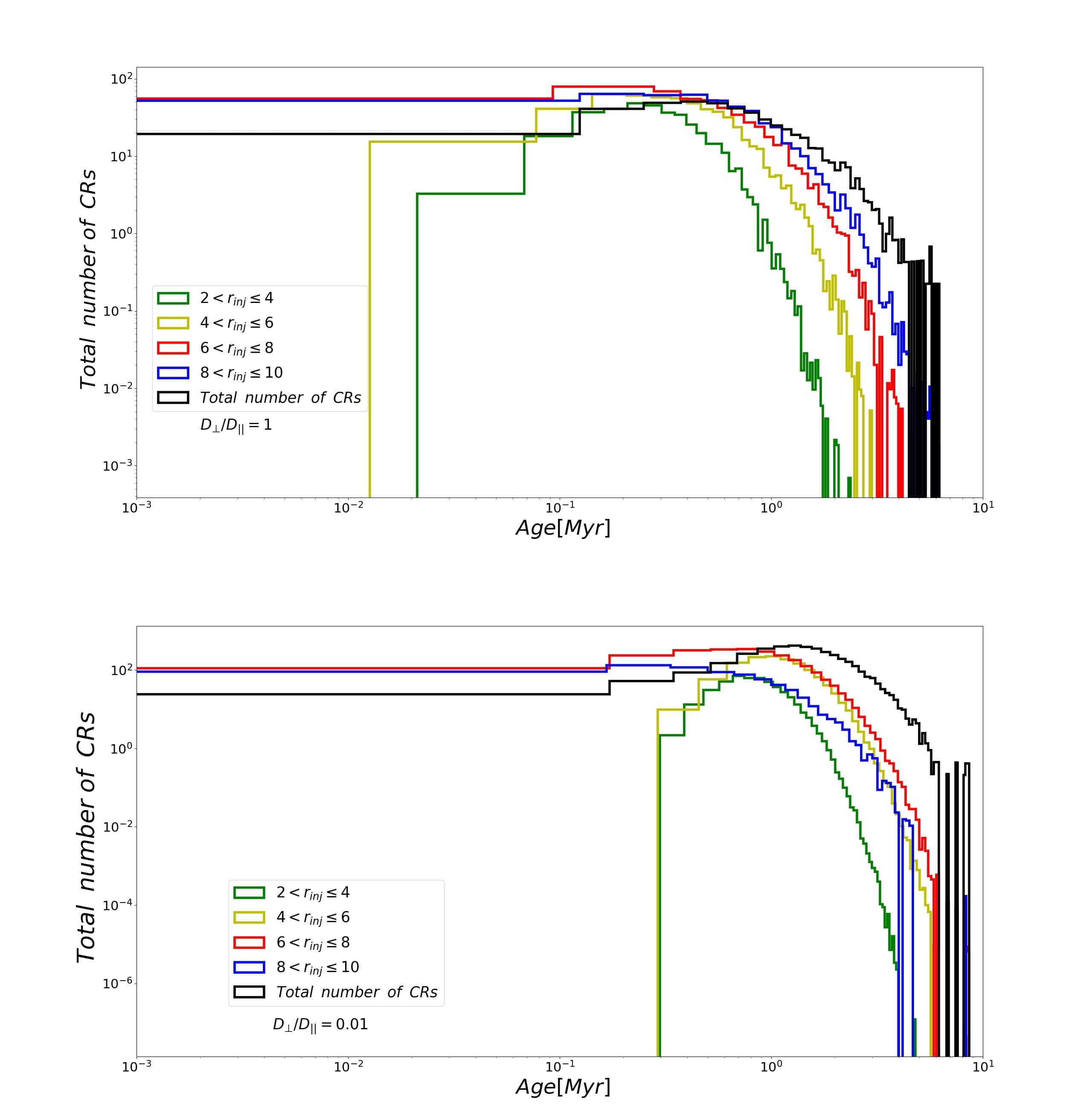}
\caption{As in figure \ref{figure4}, now using  the supernova surface density  distribution (\ref{W2}) with $R_{\rm snr} \;= 5.4 \;{\rm kpc}$. }
\label{figure7}
\end{figure} 
\clearpage
\begin{figure}
\includegraphics [width=\columnwidth]{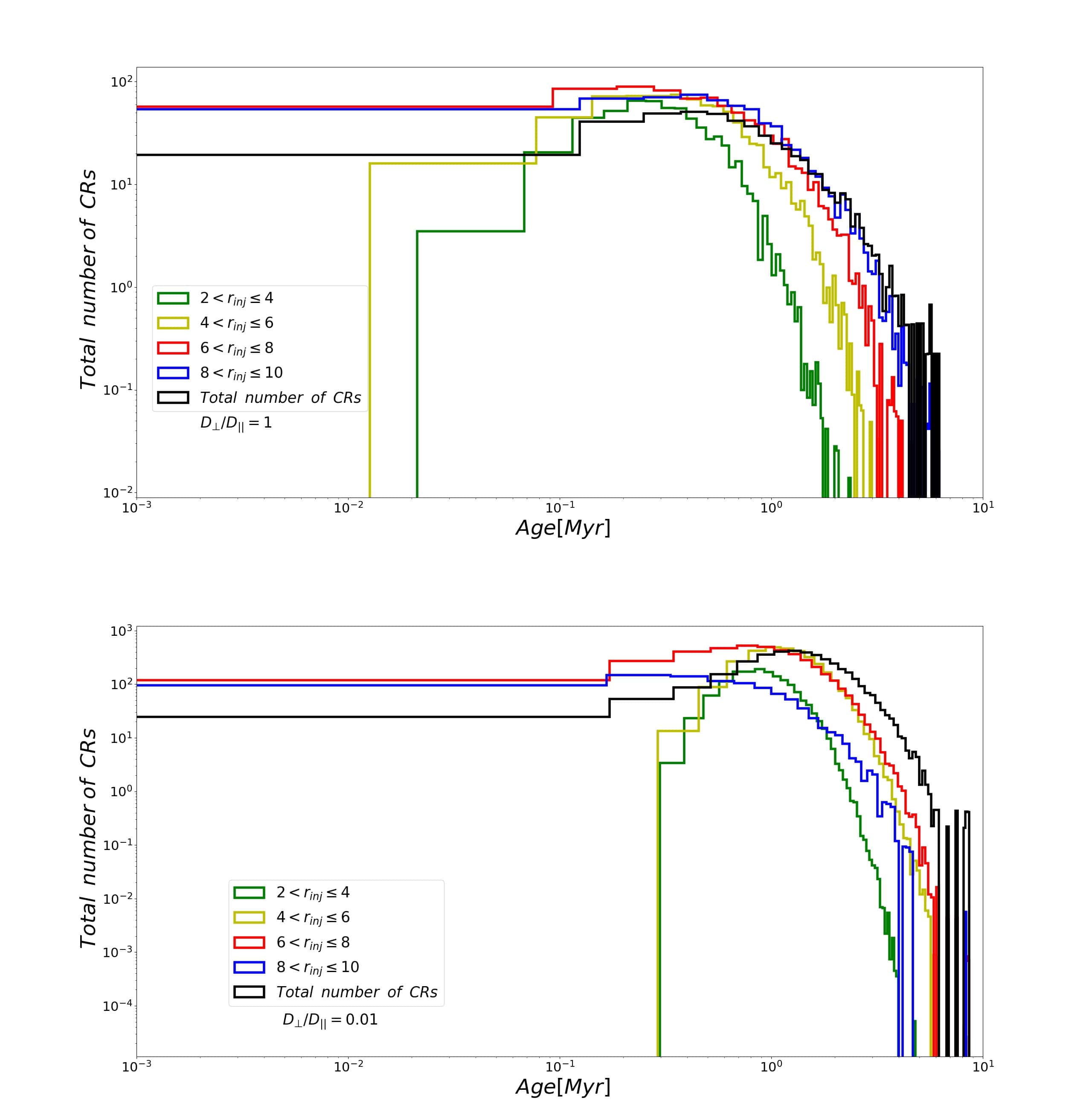}
\caption{As in figure \ref{figure8}, for a value of $R_{\rm snr} \;= 8 \;{\rm kpc}$. }
\label{figure8}
\end{figure} 
  \begin{figure}
\includegraphics [width=\columnwidth]{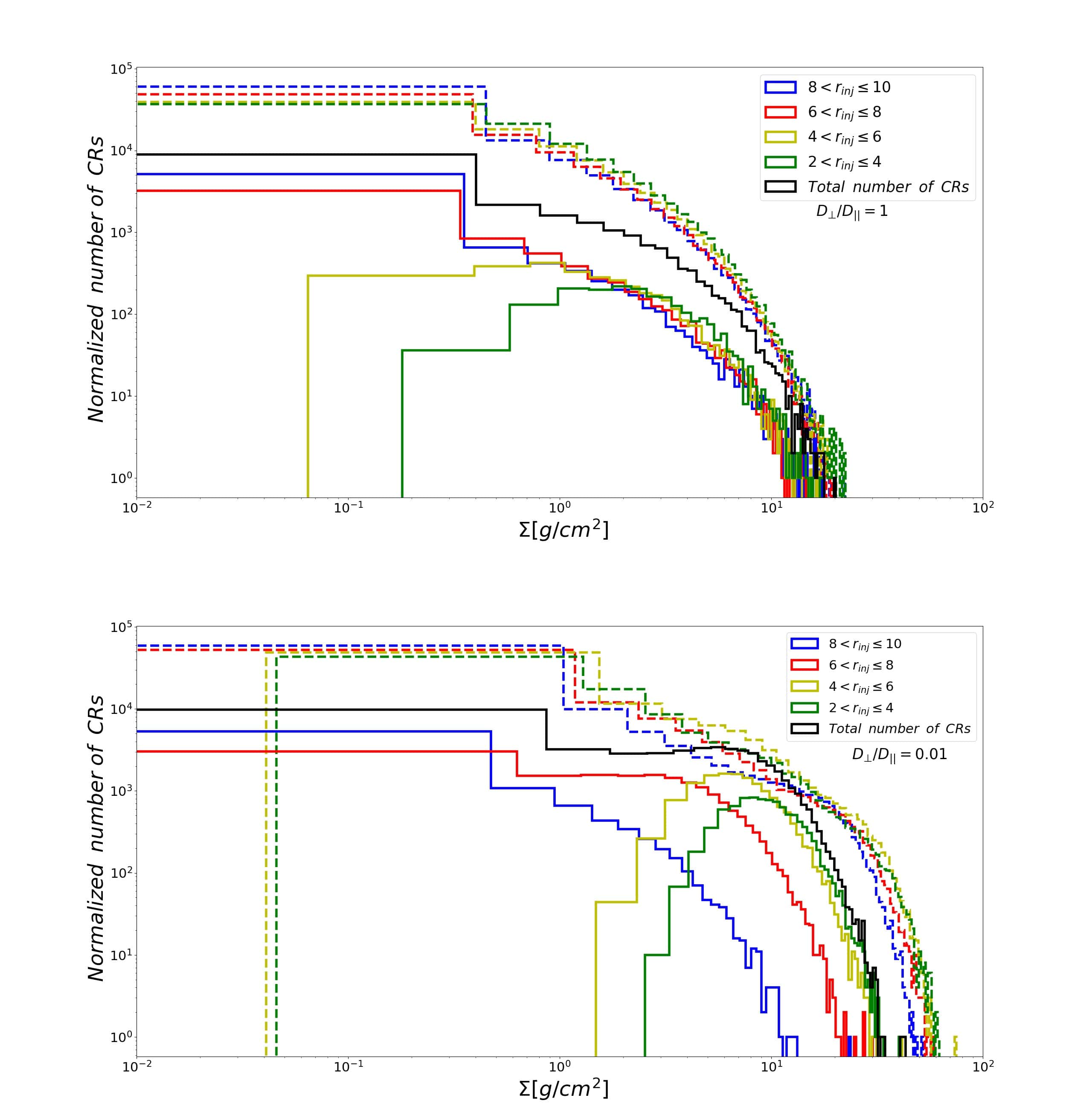}
\caption{The normalized grammage distribution for the CRs observed in the local volume around the Solar System (solid lines). 
The black solid line shows the same for all CRs.
The dashed lines show the same, but now at the moment of escape when the CRs reach the edge of the CR halo at $|z| = H_{\rm cr}$ with $H_{\rm cr} =  4 \; {\rm kpc}$. 
The two cases shown are $\Dperp/\Dpar=1$ (upper panel: isotropic diffusion) and $\Dperp/\Dpar = 0.01$ (lower panel: anisotropic diffusion).
}
\label{figure9}
\end{figure} 
 \clearpage
  \begin{figure}
  \centering
\includegraphics  [ width=14cm]{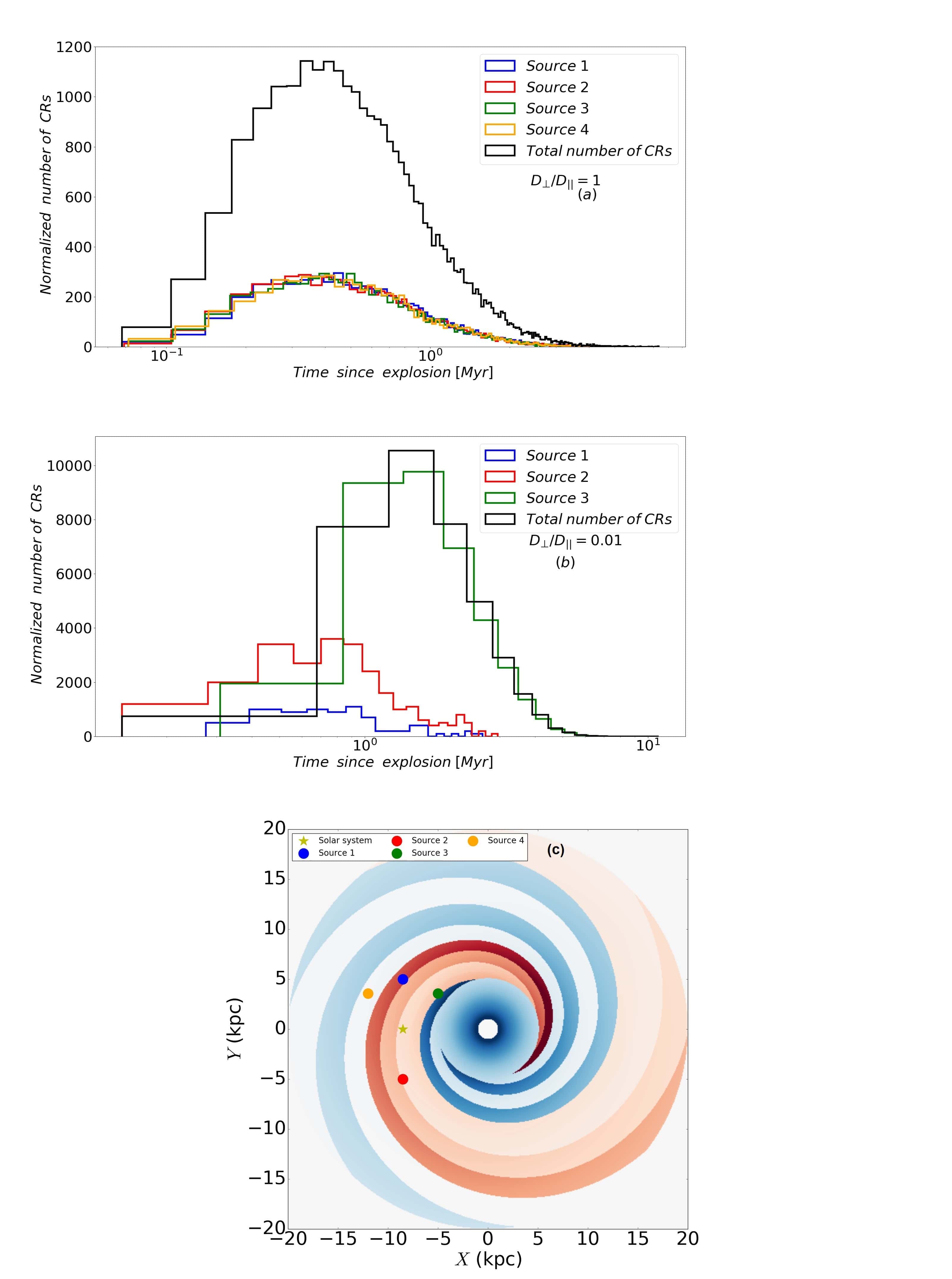}
\caption{This figure shows the CR proton flux at the position of the Sun due to a  set of four hypothetical single sources. These sources are located at a distance of 5 kpc in different sectors of the Galactic magnetic field.
The sources  marked as (1,2,3,4). The  solid black line shows the total CRs come from all four sources. For the value of the ratio $\Dperp/\Dpar=1$ (a) $\Dperp/\Dpar=0.01$(b), (c) shows the position of the discrete sources (circles) in the Galactic magnetic field. Yellow (star) indicates the position of our Solar System. }
\label{figure10}
\end{figure}

  \begin{figure}
\includegraphics [width=\columnwidth]{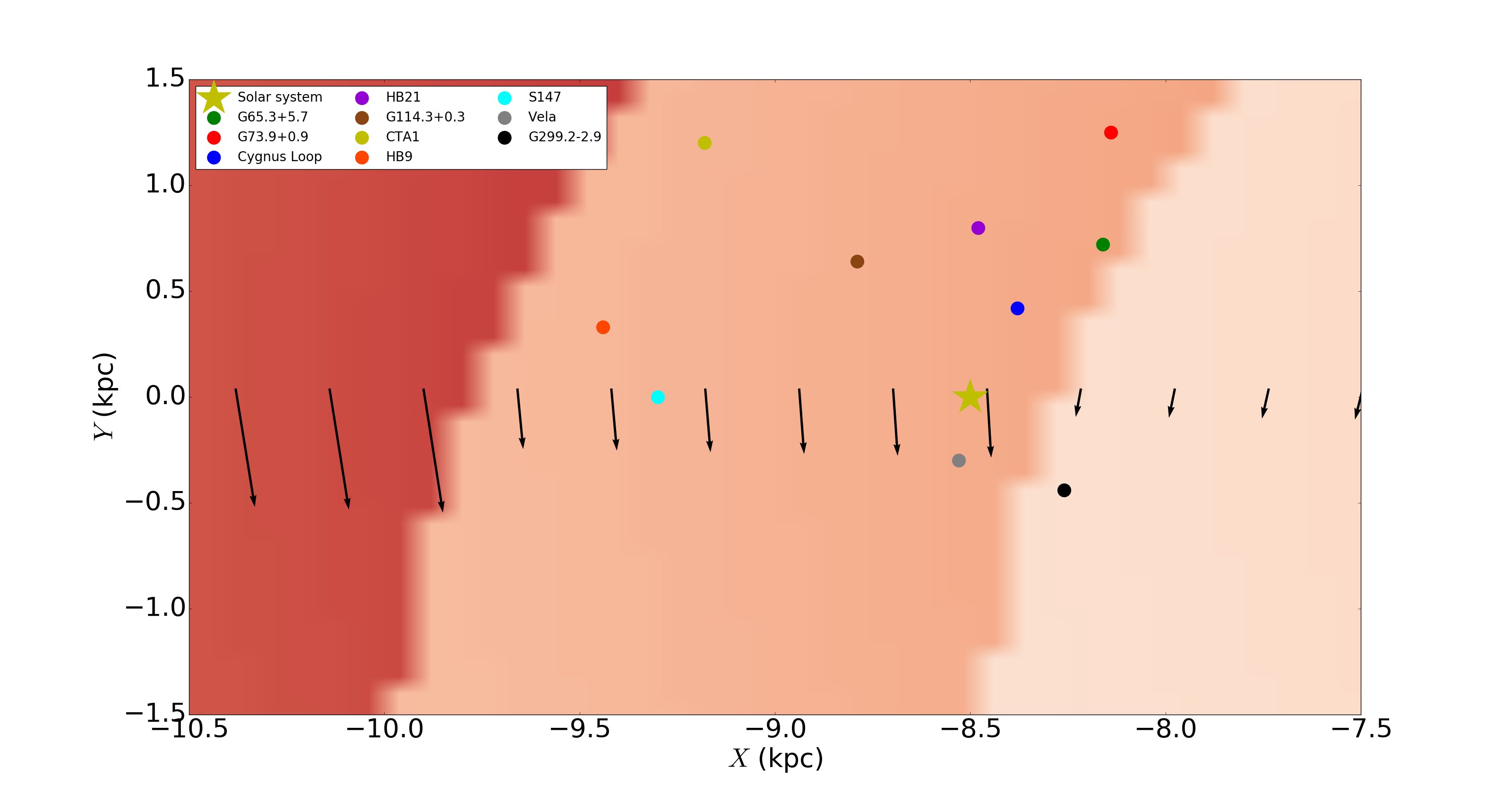}
\caption{This figure shows the position of the known nearby SNRs (circles) in the Galactic magnetic field. The yellow star indicates the position of our Solar System. 
The arrows indicate the local direction of the Galactic magnetic field, projected onto the plane of the Galactic disk.}
\label{figure11}
\end{figure} 
  \begin{figure}
\includegraphics [width=\columnwidth]{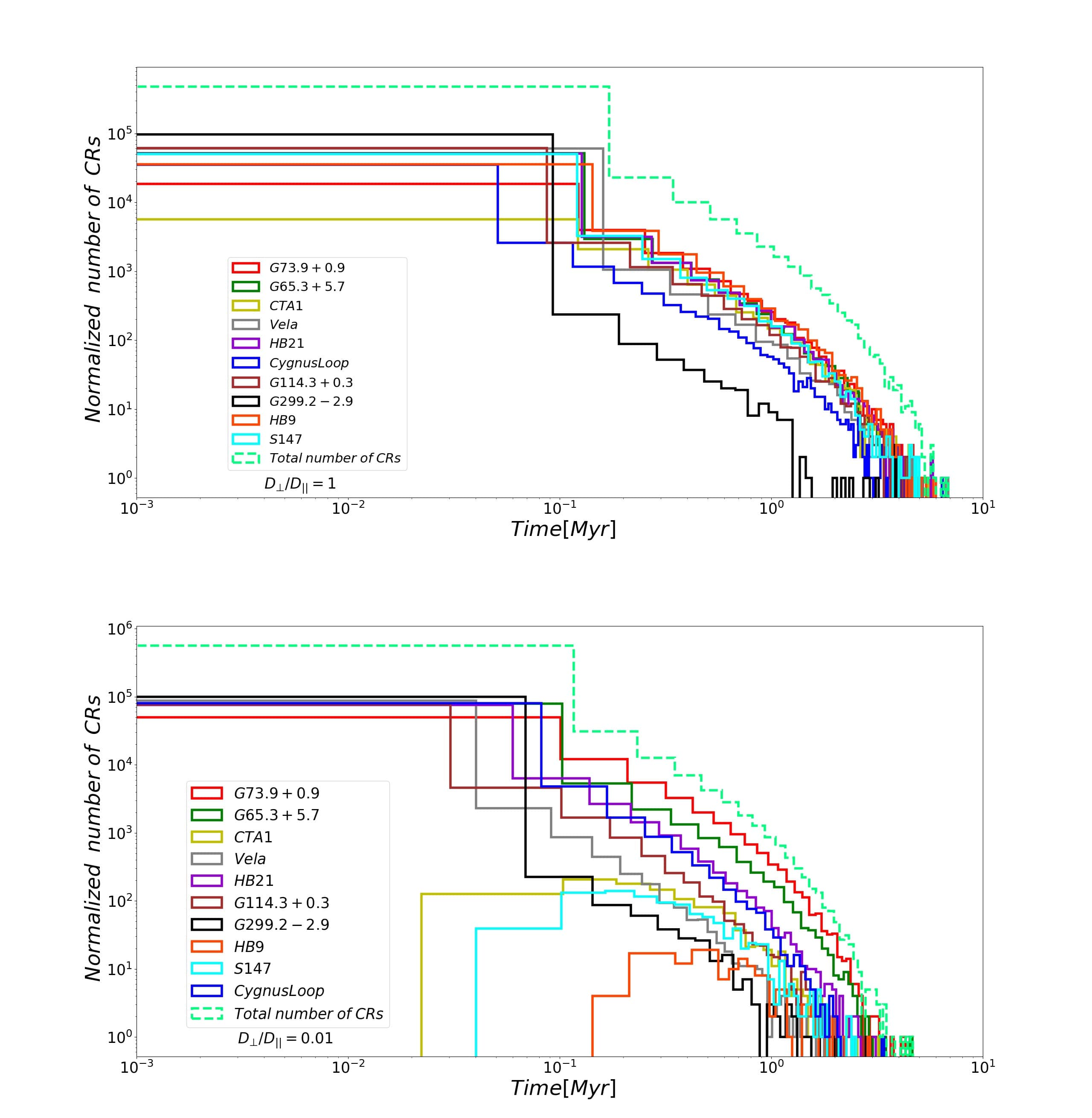}
\caption{This figure shows, as a function of time, the normalized number of CRs that reside within 0.5 kpc of the Solar System. 
The sources are the SNRs listed in table \ref{table:1}. The  dashed line shows the total number of CRs from all ten sources together.
Again we show the results for $\Dperp/\Dpar=1$ (upper panel) and $\Dperp/\Dpar=0.01$(lower panel).}
\label{figure12}
\end{figure} 
  \begin{figure}
\includegraphics [width=\columnwidth]{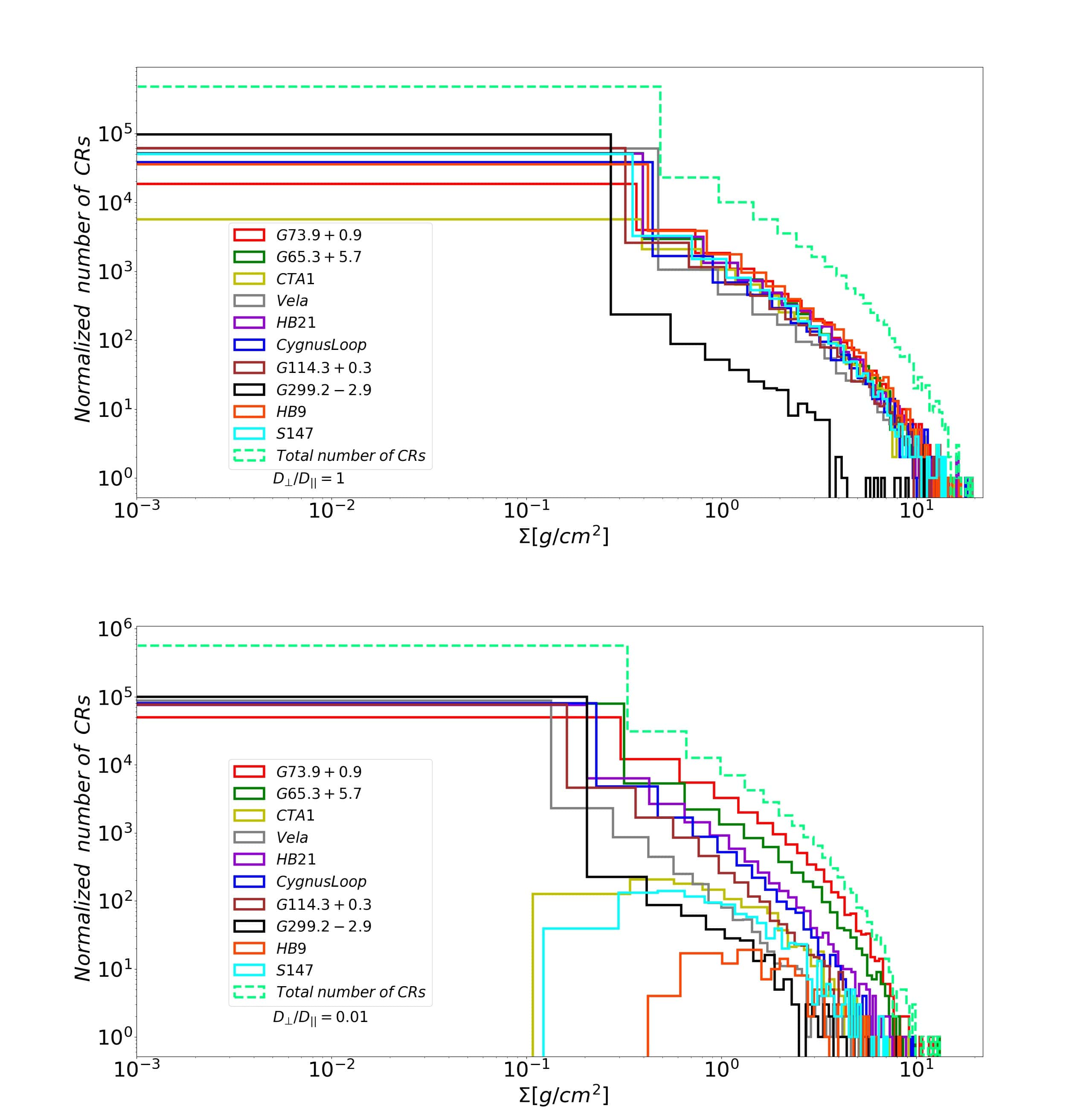}
\caption{As in figure \ref{figure12}, but now showing the distribution over grammage.}
\label{figure13}
\end{figure} 

\clearpage
 \begin{figure}
 \centering
\includegraphics [width=\columnwidth]{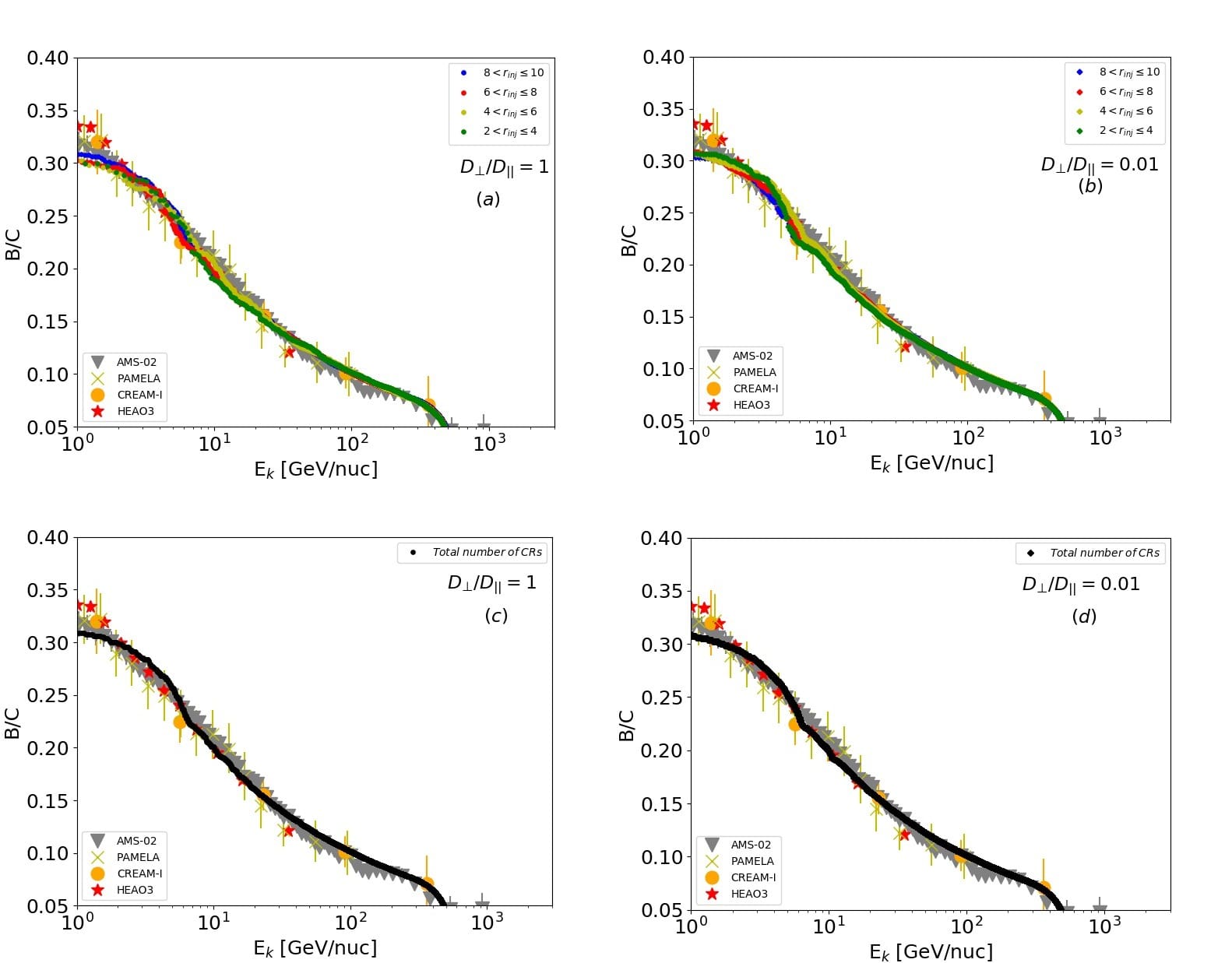}
\caption{This figure shows the Boron- Carbon  abundance ratio ( B/C) as a function of kinetic energy per nucleon for two cases of the diffusion coefficient $\Dperp/\Dpar=1 (a)\;, 0.01 (b)$  for the four rings in the local volume around the Solar System (solid lines), compared to observational data from  HEAO3 \citet{HEAO31990}, CREAM \citet{CREAM2008},
 AMS-02 \citet{AMS02-2016}, and PAMELA \citet{PAMELA2008}. The black solid line is the total number of  CRs come from all of the four rings, as indicated in each panel.  
 We use the value of $ D_{0}$ 
to produce the best fit with the observational data. We use
$ D_{0}= 3 \times 10^{28}$ cm$^{2}$ s$^{-1}$ for $\Dperp/\Dpar=0.01$ (anisotropic diffusion), and $D_{0}= 2.8 \times 10^{27}$ cm$^{2}$ s$^{-1}$ for  $ \Dperp/\Dpar=1$ ( isotropic diffusion). }
\label{figure14}
\end{figure}

\clearpage
 \begin{figure}
 \centering
\includegraphics [width=\columnwidth]{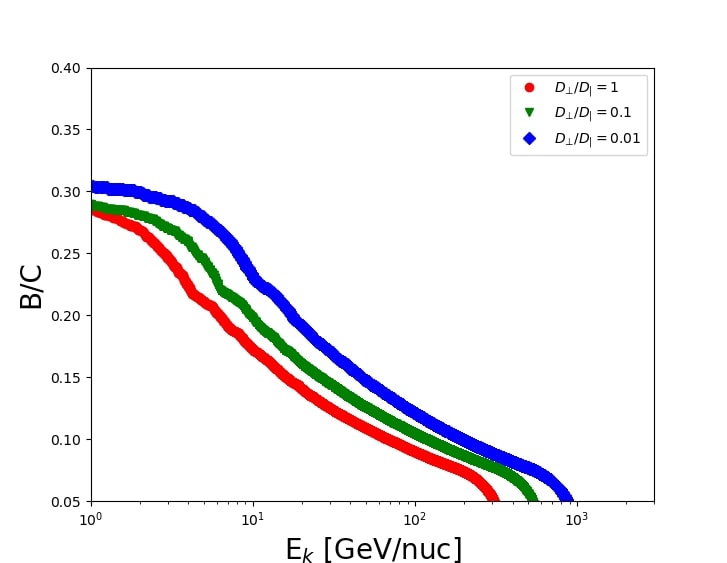}
\caption{This figure shows the Boron- Carbon   abundance ratio  (B/C) as a function of kinetic energy per nucleon for the value of the ratio $\Dperp/\Dpar= \;0.01\;,\; 0.1\;, \;1$. 
In contrast to what is shown in Figure 14 the value of $D_{\parallel}$ is now fixed at $\Dpar =4 \times 10^{28}$ cm$^{2}$ s$^{-1}$ .}
\label{figure15}
\end{figure}

\clearpage
 \begin{figure}
 \centering
\includegraphics [width=\columnwidth]{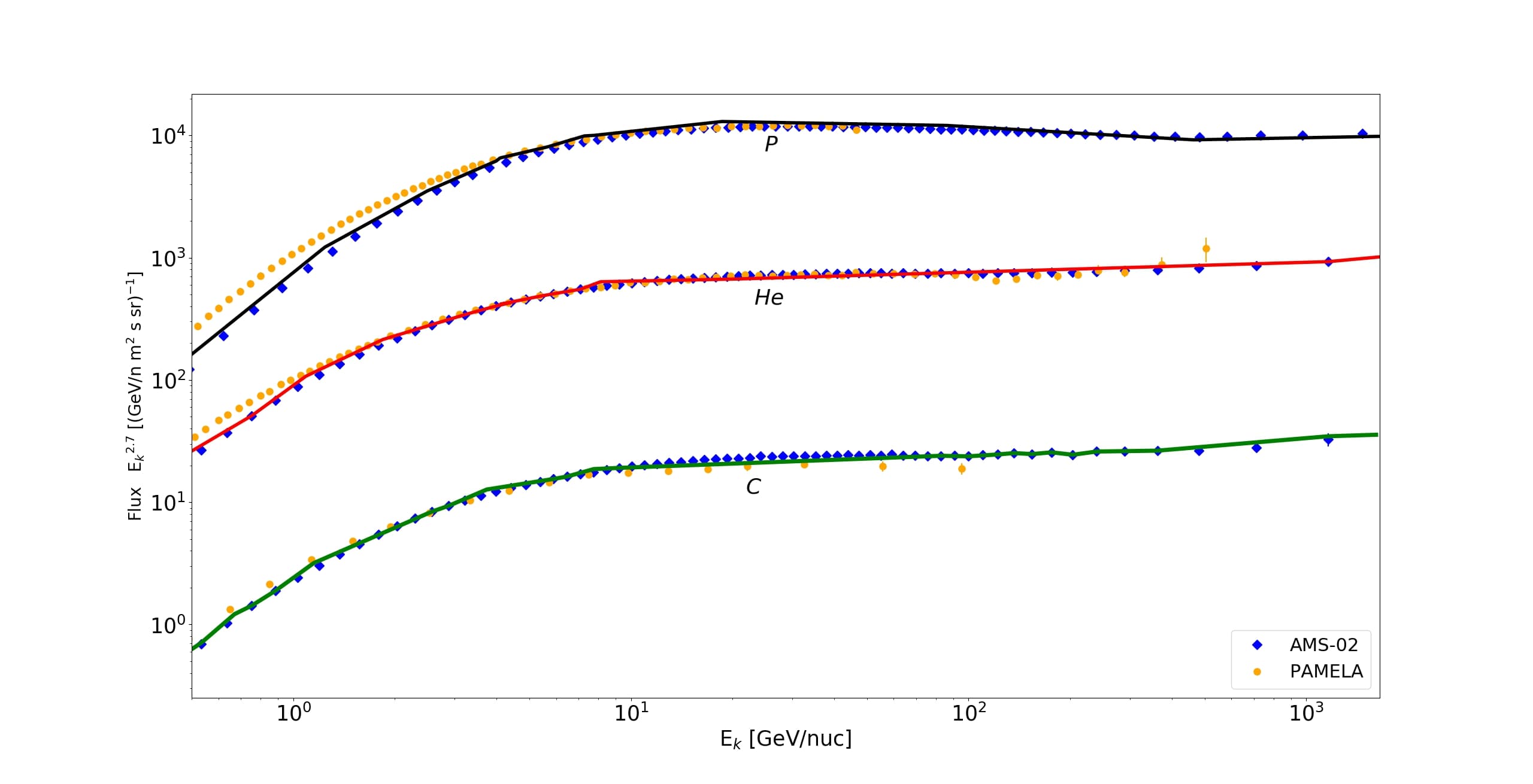}
\caption{This figure shows The spectrum of Proton, Helium and Carbon as function of kinetic energy per nucleon for the  case of the diffusion coefficient $\Dperp/\Dpar= 0.01$, compared to observational data from
 for Proton: \citet{pAMS2015} ,\citet{pPAMELA2013}, for He: \citet{CHe2017}, \citet{HePAM2011}, for C: \citet{CHe2017}, \citet{CPam2014}.}
\label{figure16}
\end{figure}
\clearpage
\clearpage
 \begin{figure}
 \centering
\includegraphics [width=\columnwidth]{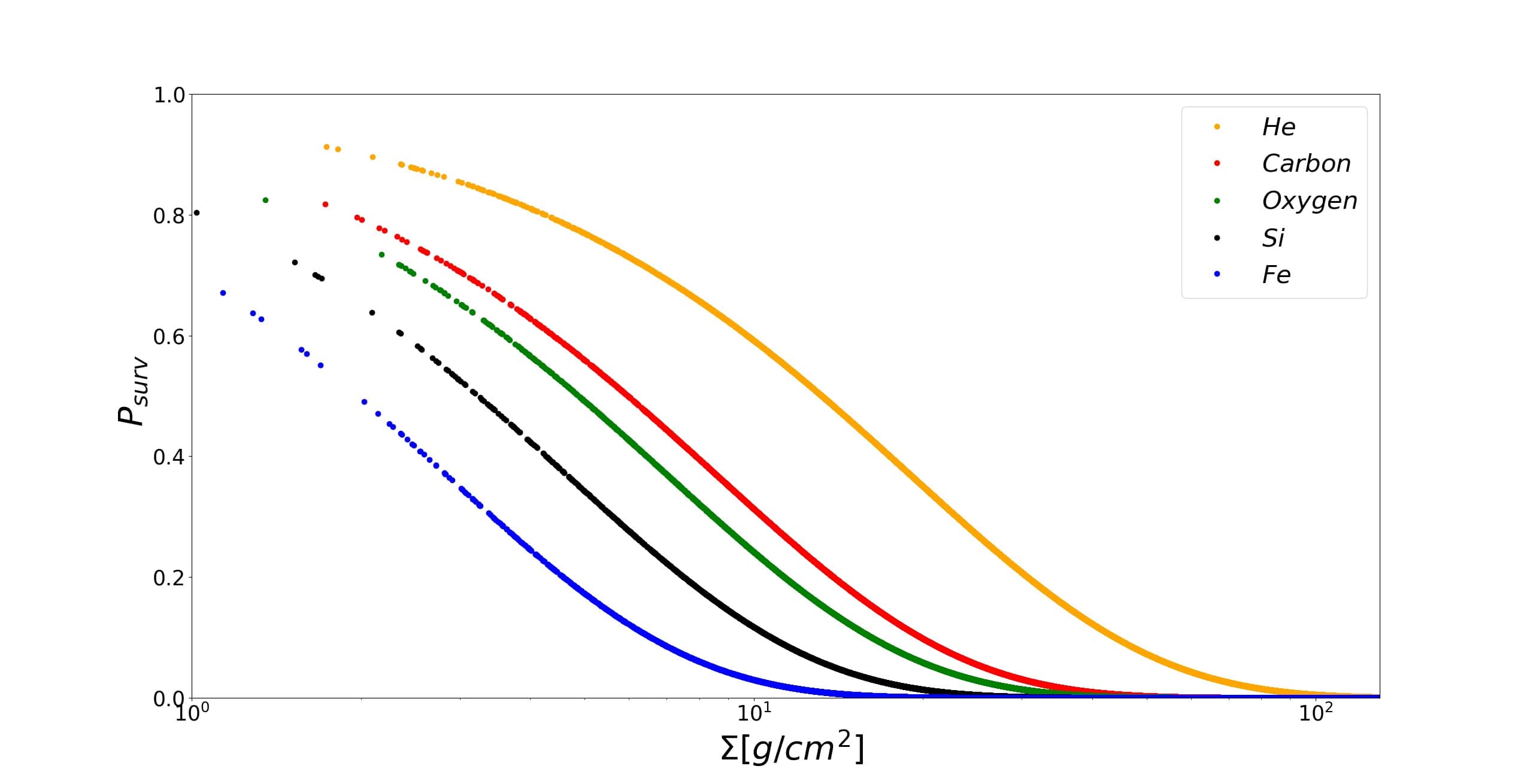}
\caption{This figure shows the survival probability for some  CR nuclei as a function of grammage.}
\label{figure17}
\end{figure}

\bibliographystyle{mnras}
\bibliography{References_Paper2}
\appendix

\section{Local and isotropic cosmic ray diffusion in the Galactic disk}

Consider cosmic rays (CRs) injected impulsively at $t = 0$, somewhere in the mid-plane of the Galactic disk. Employing a set of local cylindrical
coordinates $(\varpi \: , \: \theta \: , \: z)$ centered on the source location, with $z$
the coordinate perpendicular to the disk mid-plane $z = 0$, isotropic diffusion of the CRs with a constant diffusion coefficient $D_{\rm cr}$ is described for $t > 0$ by
the following equation for the CR density $\ncr$:
\be
	\frac{\partial \ncr}{\partial t} = \frac{D_{\rm cr}}{\varpi} \: \frac{\partial}{\partial \varpi} \left( \varpi \: \frac{\partial \ncr}{\partial \varpi} \right) +
	D_{\rm cr} \: \frac{\partial^2  \ncr}{\partial z^2} \; .
\ee 
Here we assumed axi-symmetry around the source location. This is a reasonable approximation as long as the source is not close to the edge of the Galaxy.
We assume that CRs escape instantaneously once the reach the upper/lower edge of the CR halo, located at $z = \pm \Hcr$. This leads to the condition:
\be
\label{edges}
	n_{\rm cr}(\varpi \: , \: z = \pm \Hcr \: , \: t) = 0 \; .
\ee
Solution of this equation subject to condition (\ref{edges}) proceeds by standard means: a Laplace Transformation with respect to time, 
followed by separation of the remaining independent variables $\varpi$ and $z$.
Defining the Laplace Transform of $\ncr$ as $\ncrs$ so that:
\be
\label{CRbc}
	\ncrs = \int_{0}^{\infty} {\rm d}t \: e^{-st} \: \ncr \; ,
\ee
standard theory, see for instance \citep{AW2005}, Ch. 15.8 yields:
\be
\label{LTeqn}
	\frac{1}{\varpi} \: \frac{\partial}{\partial \varpi} \left( \varpi \: \frac{\partial \ncrs}{\partial \varpi} \right) +
	 \frac{\partial^2  \ncrs}{\partial z^2} - \left( \frac{s}{D_{\rm cr}} \right) \: \ncrs = 0 \; .
\ee
We seek a solution of the form
\be
	\ncrs = R(\varpi \: , \: s) \: Z(z) \; .
\ee
In order to satisfy (\ref{CRbc}) we must demand $Z(z = \pm \Hcr) = 0$ and look for solutions where, introducing the separation constant $- k^2$,
\be
\label{Zeqn}
	\frac{1}{Z} \: \frac{{\rm d}^2 Z}{{\rm d}z^2} = - k^2 \; 
\ee
and
\be
\label{Reqn}
	\frac{1}{\varpi} \: \frac{{\rm d}}{{\rm d} \varpi} \left( \varpi \: \frac{{\rm d} R}{{\rm d} \varpi} \right) - \left( k^2 + \frac{s}{D_{\rm cr}} \right) \: R = 0 \; .
\ee	
The solutions of (\ref{Zeqn}) that satisfy the condition $Z(z = \pm \Hcr) = 0$ are:
\be
	Z(z) = \cos(k_{m}z) \; \; , \; \; k_{m} \equiv \left(m + \half \right) \: \frac{\pi}{\Hcr} \; \mbox{with $m = 0 \: , \: 1 \: , \: 2 \: , \: \cdots$} \; .
\ee
The corresponding solutions of (\ref{Reqn}) that satisfy $\ncr = 0$ for $\varpi \rightarrow \infty$ are:
\be
	R(\varpi \: , \: s) = K_{0}(q_{m}(s) \: \varpi) \; \; , \; \; q_{m}(s) \equiv \sqrt{k_{m}^2 + \frac{s}{D_{\rm cr}}} \; .
\ee
The general solution of (\ref{LTeqn}) is therefore:
\be
	\ncrs = \sum_{m = 0}^{\infty} a_{m} \: K_{0}(q_{m}(s) \: \varpi) \:  \cos(k_{m}z) \; ,
\ee
where the constant coefficients $a_{m}$ must be determined from the initial conditions. Their value need not concern us in what follows.
\nskip
The CR density $\ncr$ follows from the inverse Laplace Transform. For that we may use (c.f. \citet{AS72}, p. 1028):
\be
	K_{0}\left(a \sqrt{s + s_{0}} \right) \; \; \Longrightarrow \frac{e^{-s_{0} t}}{2t} \: {\rm exp}\left(- \frac{a^2}{4t} \right) \; .
\ee
In this particular case $a = \varpi/\sqrt{D_{\rm cr}}$ and $s_{0} = k_{m}^2 D$ and one finds:
\be
	\ncr = \frac{1}{2t} \: \sum_{m = 0}^{\infty} a_{m} \:\:  \cos(k_{m}z) \: e^{- k_{m}^2 Dt} \: {\rm exp} \left( - \frac{\varpi^2}{4 Dt} \right) \; . 
\ee
At late times, well after injection when the initial build-up of the density $\ncr$ has passed, the $m = 0$ term dominates 
(unless $a_{0} = 0$, which is generally not the case) 
because of the exponential decay factor ${\rm exp}(-k_{m}^2 D_{\rm cr}t)$. Near the mid-plane of the disk ($z = 0$) and at a distance $\varpi$ from the source 
the CR density due to this single, impulsive source then roughly scales as:
\ba
\label{ncrexpr}
	n_{\rm cr}(\varpi \: , \: 0 \: , \: t) & \propto &  \frac{1}{t} \: {\rm exp} \left[ - \left( \frac{\pi^2 D_{\rm cr}t}{4 \Hcr^2} + \frac{\varpi^2}{4 D_{\rm cr}t} \right) \right] 
	\nonumber \\
	& & \\
	& = & \frac{\pi D_{\rm cr}}{\varpi \Hcr} \: \frac{1}{\tau} \: {\rm exp} \left[ - \frac{\pi \varpi}{4 \Hcr}
	\left( \tau + \frac{1}{\tau} \right) \right] \; . \nonumber
\ea
Here the dimensionless time variable $\tau$ is defined as:
\be
	\tau = \frac{\pi \: D_{\rm cr}t}{\varpi \Hcr} \; ,
\ee
so that the argument of the exponential factor in expression (\ref{ncrexpr}) has a {\em minimum} at $\tau = 1$. If we write that factor as ${\rm exp}(- \psi)$ we have
$\psi_{\rm min} = \pi \varpi/2 \Hcr$.
The maximum value of $n_{\rm cr}(\varpi \: , \: 0 \: , \: t)$ is reached $\tau = \tau_{\rm m}$ (time $t = t_{\rm m}$) with
\be
	\tau_{\rm m}(\varpi) = \frac{2 \Hcr}{\pi \varpi} \: \left( \sqrt{\displaystyle 1 + \frac{\pi^2 \varpi^2}{4 \Hcr^2}} - 1 \right) \; \; , \; \; \; 
	t_{\rm m}(\varpi) = \frac{4 t_{\rm res}}{\pi^2} \: \left( \sqrt{\displaystyle 1 + \frac{\pi^2 \varpi^2}{4 \Hcr^2}} - 1 \right) \; . 
\ee
Here we have used $t_{\rm res} = H_{\rm cr}^2/2D_{\rm cr}$ for the typical CR residence time in the Galaxy.
Note that $\tau_{\rm m} \simeq 1$ when $\varpi \gg \Hcr$ so that the exponential factor indeed dominates temporal behavior in that limit.
Concluding: at given $\varpi$ and at late times (i.e. for $\tau \gg 1$) 
the local CR density in the mid-plane, at time $t$ after injection by an impulsive CR source at $\varpi = 0$, decays  roughly as
\be
	 n_{\rm cr}(\varpi \: , \: 0 \: , \: t= \frac{\varpi \Hcr}{\pi D_{\rm cr}} \: \tau) \propto 
	 \frac{\pi D_{\rm cr}}{\varpi \Hcr \: \tau} \: \displaystyle {\rm exp}  \left(- \frac{ \pi \varpi}{4 \Hcr} \: \tau \right) \; .
\ee
This confirms the intuitive notion that CR escape limits the spread of diffusing CRs in the disk plane to a typical distance 
$\varpi \simeq \sqrt{2D_{\rm cr} t_{\rm res}} = \Hcr$ from the source
when diffusion is isotropic.

\end{document}